\documentclass[longauth]{aa} 
\usepackage{mathptmx}
\usepackage{txfonts}
\usepackage{float}
\usepackage{capt-of}
\usepackage{multicol}
\usepackage{newtxtext}
\usepackage[T1]{fontenc}
\usepackage{lastpage}
\DeclareRobustCommand{\VAN}[3]{#2}
\let\VANthebibliography\thebibliography
\def\thebibliography{\DeclareRobustCommand{\VAN}[3]{##3}\VANthebibliography}

\usepackage{graphicx}
\usepackage[dvipsnames]{xcolor}

\usepackage{tabularx}
\usepackage[inline]{enumitem}
\usepackage{tabulary}
\usepackage{booktabs}
\usepackage{multirow}
\usepackage{placeins}
\usepackage{listings}
\usepackage{orcidlink}
\usepackage[normalem]{ulem}
\DeclareUnicodeCharacter{2248}{$\approx$}
\DeclareUnicodeCharacter{0306}{$\check{c}$}
\usepackage{amsmath}
\usepackage{xspace}
\usepackage[switch, modulo]{lineno}
\usepackage{euclid}

\makeatletter 
  \patchcmd{\NAT@citex}
    {\@citea\NAT@hyper@{%
      \NAT@nmfmt{\NAT@nm}%
      \hyper@natlinkbreak{\NAT@aysep\NAT@spacechar}{\@citeb\@extra@b@citeb}%
      \NAT@date}}
    {\@citea\NAT@nmfmt{\NAT@nm}%
    \NAT@aysep\NAT@spacechar\NAT@hyper@{\NAT@date}}{}{}

  \patchcmd{\NAT@citex}
    {\@citea\NAT@hyper@{%
      \NAT@nmfmt{\NAT@nm}%
      \hyper@natlinkbreak{\NAT@spacechar\NAT@@open\if*#1*\else#1\NAT@spacechar\fi}%
        {\@citeb\@extra@b@citeb}%
      \NAT@date}}
    {\@citea\NAT@nmfmt{\NAT@nm}%
    \NAT@spacechar\NAT@@open\if*#1*\else#1\NAT@spacechar\fi\NAT@hyper@{\NAT@date}}
    {}{}
\makeatother

\titlerunning{Peanutty dwarf galaxies in Perseus}
\title{\Euclid: Early Release Observations -- The formation of peanutty dwarf galaxies in Perseus\thanks{This paper is published on behalf of the Euclid Consortium.}}
\newcommand{\orcid}[1]{} 
\author{A.~U.~Kapoor\orcid{0000-0002-5187-1725}\thanks{\email{anandutsav.kapoor@ugent.be}}\inst{\ref{aff1}}
\and R.~Smith\orcid{0000-0001-5303-6830}\inst{\ref{aff2},\ref{aff3}}
\and A.~J.~Zuiderwijk\orcid{0009-0008-2319-3622}\inst{\ref{aff1}}
\and S.~De~Rijcke\orcid{0000-0001-7680-2059}\inst{\ref{aff1}}
\and A.~van~der~Wel\orcid{0000-0002-5027-0135}\inst{\ref{aff1}}
\and M.~Baes\orcid{0000-0002-3930-2757}\inst{\ref{aff1}}
\and F.~R.~Marleau\orcid{0000-0002-1442-2947}\inst{\ref{aff4}}
\and P.-A.~Duc\orcid{0000-0003-3343-6284}\inst{\ref{aff5}}
\and R.~F.~Peletier\orcid{0000-0001-7621-947X}\inst{\ref{aff6}}
\and B.~Altieri\orcid{0000-0003-3936-0284}\inst{\ref{aff7}}
\and S.~Andreon\orcid{0000-0002-2041-8784}\inst{\ref{aff8}}
\and N.~Auricchio\orcid{0000-0003-4444-8651}\inst{\ref{aff9}}
\and C.~Baccigalupi\orcid{0000-0002-8211-1630}\inst{\ref{aff10},\ref{aff11},\ref{aff12},\ref{aff13}}
\and M.~Baldi\orcid{0000-0003-4145-1943}\inst{\ref{aff14},\ref{aff9},\ref{aff15}}
\and A.~Balestra\orcid{0000-0002-6967-261X}\inst{\ref{aff16}}
\and S.~Bardelli\orcid{0000-0002-8900-0298}\inst{\ref{aff9}}
\and P.~Battaglia\orcid{0000-0002-7337-5909}\inst{\ref{aff9}}
\and A.~Biviano\orcid{0000-0002-0857-0732}\inst{\ref{aff11},\ref{aff10}}
\and E.~Branchini\orcid{0000-0002-0808-6908}\inst{\ref{aff17},\ref{aff18},\ref{aff8}}
\and M.~Brescia\orcid{0000-0001-9506-5680}\inst{\ref{aff19},\ref{aff20}}
\and S.~Camera\orcid{0000-0003-3399-3574}\inst{\ref{aff21},\ref{aff22},\ref{aff23}}
\and V.~Capobianco\orcid{0000-0002-3309-7692}\inst{\ref{aff23}}
\and C.~Carbone\orcid{0000-0003-0125-3563}\inst{\ref{aff24}}
\and J.~Carretero\orcid{0000-0002-3130-0204}\inst{\ref{aff25},\ref{aff26}}
\and M.~Castellano\orcid{0000-0001-9875-8263}\inst{\ref{aff27}}
\and G.~Castignani\orcid{0000-0001-6831-0687}\inst{\ref{aff9}}
\and S.~Cavuoti\orcid{0000-0002-3787-4196}\inst{\ref{aff20},\ref{aff28}}
\and K.~C.~Chambers\orcid{0000-0001-6965-7789}\inst{\ref{aff29}}
\and A.~Cimatti\inst{\ref{aff30}}
\and C.~Colodro-Conde\inst{\ref{aff31}}
\and G.~Congedo\orcid{0000-0003-2508-0046}\inst{\ref{aff32}}
\and C.~J.~Conselice\orcid{0000-0003-1949-7638}\inst{\ref{aff33}}
\and L.~Conversi\orcid{0000-0002-6710-8476}\inst{\ref{aff34},\ref{aff7}}
\and Y.~Copin\orcid{0000-0002-5317-7518}\inst{\ref{aff35}}
\and F.~Courbin\orcid{0000-0003-0758-6510}\inst{\ref{aff36},\ref{aff37},\ref{aff38}}
\and H.~M.~Courtois\orcid{0000-0003-0509-1776}\inst{\ref{aff39}}
\and M.~Cropper\orcid{0000-0003-4571-9468}\inst{\ref{aff40}}
\and H.~Degaudenzi\orcid{0000-0002-5887-6799}\inst{\ref{aff41}}
\and G.~De~Lucia\orcid{0000-0002-6220-9104}\inst{\ref{aff11}}
\and H.~Dole\orcid{0000-0002-9767-3839}\inst{\ref{aff42}}
\and F.~Dubath\orcid{0000-0002-6533-2810}\inst{\ref{aff41}}
\and X.~Dupac\inst{\ref{aff7}}
\and M.~Fabricius\orcid{0000-0002-7025-6058}\inst{\ref{aff43},\ref{aff44}}
\and M.~Farina\orcid{0000-0002-3089-7846}\inst{\ref{aff45}}
\and R.~Farinelli\inst{\ref{aff9}}
\and S.~Ferriol\inst{\ref{aff35}}
\and M.~Frailis\orcid{0000-0002-7400-2135}\inst{\ref{aff11}}
\and S.~Galeotta\orcid{0000-0002-3748-5115}\inst{\ref{aff11}}
\and K.~George\orcid{0000-0002-1734-8455}\inst{\ref{aff46}}
\and B.~Gillis\orcid{0000-0002-4478-1270}\inst{\ref{aff32}}
\and C.~Giocoli\orcid{0000-0002-9590-7961}\inst{\ref{aff9},\ref{aff15}}
\and J.~Gracia-Carpio\orcid{0000-0003-4689-3134}\inst{\ref{aff43}}
\and A.~Grazian\orcid{0000-0002-5688-0663}\inst{\ref{aff16}}
\and F.~Grupp\inst{\ref{aff43},\ref{aff44}}
\and S.~V.~H.~Haugan\orcid{0000-0001-9648-7260}\inst{\ref{aff47}}
\and J.~Hoar\inst{\ref{aff7}}
\and W.~Holmes\orcid{0009-0007-8554-4646}\inst{\ref{aff48}}
\and I.~M.~Hook\orcid{0000-0002-2960-978X}\inst{\ref{aff49}}
\and F.~Hormuth\inst{\ref{aff50}}
\and A.~Hornstrup\orcid{0000-0002-3363-0936}\inst{\ref{aff51},\ref{aff52}}
\and K.~Jahnke\orcid{0000-0003-3804-2137}\inst{\ref{aff53}}
\and M.~Jhabvala\inst{\ref{aff54}}
\and S.~Kermiche\orcid{0000-0002-0302-5735}\inst{\ref{aff55}}
\and A.~Kiessling\orcid{0000-0002-2590-1273}\inst{\ref{aff48}}
\and B.~Kubik\orcid{0009-0006-5823-4880}\inst{\ref{aff35}}
\and M.~K\"ummel\orcid{0000-0003-2791-2117}\inst{\ref{aff44}}
\and M.~Kunz\orcid{0000-0002-3052-7394}\inst{\ref{aff56}}
\and H.~Kurki-Suonio\orcid{0000-0002-4618-3063}\inst{\ref{aff57},\ref{aff58}}
\and A.~M.~C.~Le~Brun\orcid{0000-0002-0936-4594}\inst{\ref{aff59}}
\and S.~Ligori\orcid{0000-0003-4172-4606}\inst{\ref{aff23}}
\and P.~B.~Lilje\orcid{0000-0003-4324-7794}\inst{\ref{aff47}}
\and V.~Lindholm\orcid{0000-0003-2317-5471}\inst{\ref{aff57},\ref{aff58}}
\and I.~Lloro\orcid{0000-0001-5966-1434}\inst{\ref{aff60}}
\and G.~Mainetti\orcid{0000-0003-2384-2377}\inst{\ref{aff61}}
\and O.~Mansutti\orcid{0000-0001-5758-4658}\inst{\ref{aff11}}
\and O.~Marggraf\orcid{0000-0001-7242-3852}\inst{\ref{aff62}}
\and M.~Martinelli\orcid{0000-0002-6943-7732}\inst{\ref{aff27},\ref{aff63}}
\and N.~Martinet\orcid{0000-0003-2786-7790}\inst{\ref{aff64}}
\and F.~Marulli\orcid{0000-0002-8850-0303}\inst{\ref{aff65},\ref{aff9},\ref{aff15}}
\and R.~J.~Massey\orcid{0000-0002-6085-3780}\inst{\ref{aff66}}
\and E.~Medinaceli\orcid{0000-0002-4040-7783}\inst{\ref{aff9}}
\and S.~Mei\orcid{0000-0002-2849-559X}\inst{\ref{aff67},\ref{aff68}}
\and M.~Meneghetti\orcid{0000-0003-1225-7084}\inst{\ref{aff9},\ref{aff15}}
\and E.~Merlin\orcid{0000-0001-6870-8900}\inst{\ref{aff16}}
\and G.~Meylan\orcid{0000-0001-6503-0209}\inst{\ref{aff69}}
\and A.~Mora\orcid{0000-0002-1922-8529}\inst{\ref{aff70}}
\and M.~Moresco\orcid{0000-0002-7616-7136}\inst{\ref{aff65},\ref{aff9}}
\and L.~Moscardini\orcid{0000-0002-3473-6716}\inst{\ref{aff65},\ref{aff9},\ref{aff15}}
\and R.~Nakajima\orcid{0009-0009-1213-7040}\inst{\ref{aff62}}
\and C.~Neissner\orcid{0000-0001-8524-4968}\inst{\ref{aff71},\ref{aff26}}
\and S.-M.~Niemi\orcid{0009-0005-0247-0086}\inst{\ref{aff72}}
\and C.~Padilla\orcid{0000-0001-7951-0166}\inst{\ref{aff71}}
\and S.~Paltani\orcid{0000-0002-8108-9179}\inst{\ref{aff41}}
\and F.~Pasian\orcid{0000-0002-4869-3227}\inst{\ref{aff11}}
\and K.~Pedersen\inst{\ref{aff73}}
\and W.~J.~Percival\orcid{0000-0002-0644-5727}\inst{\ref{aff74},\ref{aff75},\ref{aff76}}
\and V.~Pettorino\orcid{0000-0002-4203-9320}\inst{\ref{aff72}}
\and G.~Polenta\orcid{0000-0003-4067-9196}\inst{\ref{aff77}}
\and M.~Poncet\inst{\ref{aff78}}
\and L.~A.~Popa\inst{\ref{aff79}}
\and F.~Raison\orcid{0000-0002-7819-6918}\inst{\ref{aff43}}
\and A.~Renzi\orcid{0000-0001-9856-1970}\inst{\ref{aff80},\ref{aff81},\ref{aff9}}
\and J.~Rhodes\orcid{0000-0002-4485-8549}\inst{\ref{aff48}}
\and G.~Riccio\inst{\ref{aff20}}
\and E.~Romelli\orcid{0000-0003-3069-9222}\inst{\ref{aff11}}
\and M.~Roncarelli\orcid{0000-0001-9587-7822}\inst{\ref{aff9}}
\and R.~Saglia\orcid{0000-0003-0378-7032}\inst{\ref{aff44},\ref{aff43}}
\and Z.~Sakr\orcid{0000-0002-4823-3757}\inst{\ref{aff82},\ref{aff83},\ref{aff84}}
\and D.~Sapone\orcid{0000-0001-7089-4503}\inst{\ref{aff85}}
\and P.~Schneider\orcid{0000-0001-8561-2679}\inst{\ref{aff62}}
\and A.~Secroun\orcid{0000-0003-0505-3710}\inst{\ref{aff55}}
\and E.~Sihvola\orcid{0000-0003-1804-7715}\inst{\ref{aff86}}
\and P.~Simon\inst{\ref{aff62}}
\and C.~Sirignano\orcid{0000-0002-0995-7146}\inst{\ref{aff80},\ref{aff81}}
\and G.~Sirri\orcid{0000-0003-2626-2853}\inst{\ref{aff15}}
\and L.~Stanco\orcid{0000-0002-9706-5104}\inst{\ref{aff81}}
\and P.~Tallada-Cresp\'{i}\orcid{0000-0002-1336-8328}\inst{\ref{aff25},\ref{aff26}}
\and A.~N.~Taylor\inst{\ref{aff32}}
\and I.~Tereno\orcid{0000-0002-4537-6218}\inst{\ref{aff87},\ref{aff88}}
\and S.~Toft\orcid{0000-0003-3631-7176}\inst{\ref{aff89},\ref{aff90}}
\and R.~Toledo-Moreo\orcid{0000-0002-2997-4859}\inst{\ref{aff91},\ref{aff92}}
\and F.~Torradeflot\orcid{0000-0003-1160-1517}\inst{\ref{aff26},\ref{aff25}}
\and I.~Tutusaus\orcid{0000-0002-3199-0399}\inst{\ref{aff93},\ref{aff94},\ref{aff83}}
\and J.~Valiviita\orcid{0000-0001-6225-3693}\inst{\ref{aff57},\ref{aff58}}
\and T.~Vassallo\orcid{0000-0001-6512-6358}\inst{\ref{aff11},\ref{aff46}}
\and A.~Veropalumbo\orcid{0000-0003-2387-1194}\inst{\ref{aff8},\ref{aff18},\ref{aff17}}
\and Y.~Wang\orcid{0000-0002-4749-2984}\inst{\ref{aff95}}
\and J.~Weller\orcid{0000-0002-8282-2010}\inst{\ref{aff44},\ref{aff43}}
\and G.~Zamorani\orcid{0000-0002-2318-301X}\inst{\ref{aff9}}
\and F.~M.~Zerbi\orcid{0000-0002-9996-973X}\inst{\ref{aff8}}
\and I.~A.~Zinchenko\orcid{0000-0002-2944-2449}\inst{\ref{aff96}}
\and E.~Zucca\orcid{0000-0002-5845-8132}\inst{\ref{aff9}}
\and M.~Magliocchetti\orcid{0000-0001-9158-4838}\inst{\ref{aff45}}
\and M.~Sereno\orcid{0000-0003-0302-0325}\inst{\ref{aff9},\ref{aff15}}
\and F.~Durret\orcid{0000-0002-6991-4578}\inst{\ref{aff97}}
\and D.~Scott\orcid{0000-0002-6878-9840}\inst{\ref{aff98}}}
										   
\institute{Sterrenkundig Observatorium, Universiteit Gent, Krijgslaan 281 S9, 9000 Gent, Belgium\label{aff1}
\and
Departamento de F\'isica, Universidad Tecnica Federico Santa Maria, Avenida Espa\~na 1680, Valpara\'iso, Chile\label{aff2}
\and
Millennium Nucleus for Galaxies (MINGAL), Valpara\'iso, Chile\label{aff3}
\and
Universit\"at Innsbruck, Institut f\"ur Astro- und Teilchenphysik, Technikerstr. 25/8, 6020 Innsbruck, Austria\label{aff4}
\and
Universit\'e de Strasbourg, CNRS, Observatoire astronomique de Strasbourg, UMR 7550, 67000 Strasbourg, France\label{aff5}
\and
Kapteyn Astronomical Institute, University of Groningen, PO Box 800, 9700 AV Groningen, The Netherlands\label{aff6}
\and
ESAC/ESA, Camino Bajo del Castillo, s/n., Urb. Villafranca del Castillo, 28692 Villanueva de la Ca\~nada, Madrid, Spain\label{aff7}
\and
INAF-Osservatorio Astronomico di Brera, Via Brera 28, 20122 Milano, Italy\label{aff8}
\and
INAF-Osservatorio di Astrofisica e Scienza dello Spazio di Bologna, Via Piero Gobetti 93/3, 40129 Bologna, Italy\label{aff9}
\and
IFPU, Institute for Fundamental Physics of the Universe, via Beirut 2, 34151 Trieste, Italy\label{aff10}
\and
INAF-Osservatorio Astronomico di Trieste, Via G. B. Tiepolo 11, 34143 Trieste, Italy\label{aff11}
\and
INFN, Sezione di Trieste, Via Valerio 2, 34127 Trieste TS, Italy\label{aff12}
\and
SISSA, International School for Advanced Studies, Via Bonomea 265, 34136 Trieste TS, Italy\label{aff13}
\and
Dipartimento di Fisica e Astronomia, Universit\`a di Bologna, Via Gobetti 93/2, 40129 Bologna, Italy\label{aff14}
\and
INFN-Sezione di Bologna, Viale Berti Pichat 6/2, 40127 Bologna, Italy\label{aff15}
\and
INAF-Osservatorio Astronomico di Padova, Via dell'Osservatorio 5, 35122 Padova, Italy\label{aff16}
\and
Dipartimento di Fisica, Universit\`a di Genova, Via Dodecaneso 33, 16146, Genova, Italy\label{aff17}
\and
INFN-Sezione di Genova, Via Dodecaneso 33, 16146, Genova, Italy\label{aff18}
\and
Department of Physics "E. Pancini", University Federico II, Via Cinthia 6, 80126, Napoli, Italy\label{aff19}
\and
INAF-Osservatorio Astronomico di Capodimonte, Via Moiariello 16, 80131 Napoli, Italy\label{aff20}
\and
Dipartimento di Fisica, Universit\`a degli Studi di Torino, Via P. Giuria 1, 10125 Torino, Italy\label{aff21}
\and
INFN-Sezione di Torino, Via P. Giuria 1, 10125 Torino, Italy\label{aff22}
\and
INAF-Osservatorio Astrofisico di Torino, Via Osservatorio 20, 10025 Pino Torinese (TO), Italy\label{aff23}
\and
INAF-IASF Milano, Via Alfonso Corti 12, 20133 Milano, Italy\label{aff24}
\and
Centro de Investigaciones Energ\'eticas, Medioambientales y Tecnol\'ogicas (CIEMAT), Avenida Complutense 40, 28040 Madrid, Spain\label{aff25}
\and
Port d'Informaci\'{o} Cient\'{i}fica, Campus UAB, C. Albareda s/n, 08193 Bellaterra (Barcelona), Spain\label{aff26}
\and
INAF-Osservatorio Astronomico di Roma, Via Frascati 33, 00078 Monteporzio Catone, Italy\label{aff27}
\and
INFN section of Naples, Via Cinthia 6, 80126, Napoli, Italy\label{aff28}
\and
Institute for Astronomy, University of Hawaii, 2680 Woodlawn Drive, Honolulu, HI 96822, USA\label{aff29}
\and
Dipartimento di Fisica e Astronomia "Augusto Righi" - Alma Mater Studiorum Universit\`a di Bologna, Viale Berti Pichat 6/2, 40127 Bologna, Italy\label{aff30}
\and
Instituto de Astrof\'{\i}sica de Canarias, E-38205 La Laguna, Tenerife, Spain\label{aff31}
\and
Institute for Astronomy, University of Edinburgh, Royal Observatory, Blackford Hill, Edinburgh EH9 3HJ, UK\label{aff32}
\and
Jodrell Bank Centre for Astrophysics, Department of Physics and Astronomy, University of Manchester, Oxford Road, Manchester M13 9PL, UK\label{aff33}
\and
European Space Agency/ESRIN, Largo Galileo Galilei 1, 00044 Frascati, Roma, Italy\label{aff34}
\and
Universit\'e Claude Bernard Lyon 1, CNRS/IN2P3, IP2I Lyon, UMR 5822, Villeurbanne, F-69100, France\label{aff35}
\and
Institut de Ci\`{e}ncies del Cosmos (ICCUB), Universitat de Barcelona (IEEC-UB), Mart\'{i} i Franqu\`{e}s 1, 08028 Barcelona, Spain\label{aff36}
\and
Instituci\'o Catalana de Recerca i Estudis Avan\c{c}ats (ICREA), Passeig de Llu\'{\i}s Companys 23, 08010 Barcelona, Spain\label{aff37}
\and
Institut de Ciencies de l'Espai (IEEC-CSIC), Campus UAB, Carrer de Can Magrans, s/n Cerdanyola del Vall\'es, 08193 Barcelona, Spain\label{aff38}
\and
UCB Lyon 1, CNRS/IN2P3, IUF, IP2I Lyon, 4 rue Enrico Fermi, 69622 Villeurbanne, France\label{aff39}
\and
Mullard Space Science Laboratory, University College London, Holmbury St Mary, Dorking, Surrey RH5 6NT, UK\label{aff40}
\and
Department of Astronomy, University of Geneva, ch. d'Ecogia 16, 1290 Versoix, Switzerland\label{aff41}
\and
Universit\'e Paris-Saclay, CNRS, Institut d'astrophysique spatiale, 91405, Orsay, France\label{aff42}
\and
Max Planck Institute for Extraterrestrial Physics, Giessenbachstr. 1, 85748 Garching, Germany\label{aff43}
\and
Universit\"ats-Sternwarte M\"unchen, Fakult\"at f\"ur Physik, Ludwig-Maximilians-Universit\"at M\"unchen, Scheinerstr.~1, 81679 M\"unchen, Germany\label{aff44}
\and
INAF-Istituto di Astrofisica e Planetologia Spaziali, via del Fosso del Cavaliere, 100, 00100 Roma, Italy\label{aff45}
\and
University Observatory, LMU Faculty of Physics, Scheinerstr.~1, 81679 Munich, Germany\label{aff46}
\and
Institute of Theoretical Astrophysics, University of Oslo, P.O. Box 1029 Blindern, 0315 Oslo, Norway\label{aff47}
\and
Jet Propulsion Laboratory, California Institute of Technology, 4800 Oak Grove Drive, Pasadena, CA, 91109, USA\label{aff48}
\and
Department of Physics, Lancaster University, Lancaster, LA1 4YB, UK\label{aff49}
\and
Felix Hormuth Engineering, Goethestr. 17, 69181 Leimen, Germany\label{aff50}
\and
Technical University of Denmark, Elektrovej 327, 2800 Kgs. Lyngby, Denmark\label{aff51}
\and
Cosmic Dawn Center (DAWN), Denmark\label{aff52}
\and
Max-Planck-Institut f\"ur Astronomie, K\"onigstuhl 17, 69117 Heidelberg, Germany\label{aff53}
\and
NASA Goddard Space Flight Center, Greenbelt, MD 20771, USA\label{aff54}
\and
Aix-Marseille Universit\'e, CNRS/IN2P3, CPPM, Marseille, France\label{aff55}
\and
Universit\'e de Gen\`eve, D\'epartement de Physique Th\'eorique and Centre for Astroparticle Physics, 24 quai Ernest-Ansermet, CH-1211 Gen\`eve 4, Switzerland\label{aff56}
\and
Department of Physics, P.O. Box 64, University of Helsinki, 00014 Helsinki, Finland\label{aff57}
\and
Helsinki Institute of Physics, Gustaf H{\"a}llstr{\"o}min katu 2, University of Helsinki, 00014 Helsinki, Finland\label{aff58}
\and
Laboratoire d'etude de l'Univers et des phenomenes eXtremes, Observatoire de Paris, Universit\'e PSL, Sorbonne Universit\'e, CNRS, 92190 Meudon, France\label{aff59}
\and
SKAO, Jodrell Bank, Lower Withington, Macclesfield SK11 9FT, UK\label{aff60}
\and
Centre de Calcul de l'IN2P3/CNRS, 21 avenue Pierre de Coubertin 69627 Villeurbanne Cedex, France\label{aff61}
\and
Universit\"at Bonn, Argelander-Institut f\"ur Astronomie, Auf dem H\"ugel 71, 53121 Bonn, Germany\label{aff62}
\and
INFN-Sezione di Roma, Piazzale Aldo Moro, 2 - c/o Dipartimento di Fisica, Edificio G. Marconi, 00185 Roma, Italy\label{aff63}
\and
Aix-Marseille Universit\'e, CNRS, CNES, LAM, Marseille, France\label{aff64}
\and
Dipartimento di Fisica e Astronomia "Augusto Righi" - Alma Mater Studiorum Universit\`a di Bologna, via Piero Gobetti 93/2, 40129 Bologna, Italy\label{aff65}
\and
Department of Physics, Institute for Computational Cosmology, Durham University, South Road, Durham, DH1 3LE, UK\label{aff66}
\and
Universit\'e Paris Cit\'e, CNRS, Astroparticule et Cosmologie, 75013 Paris, France\label{aff67}
\and
CNRS-UCB International Research Laboratory, Centre Pierre Bin\'etruy, IRL2007, CPB-IN2P3, Berkeley, USA\label{aff68}
\and
Institute of Physics, Laboratory of Astrophysics, Ecole Polytechnique F\'ed\'erale de Lausanne (EPFL), Observatoire de Sauverny, 1290 Versoix, Switzerland\label{aff69}
\and
Telespazio UK S.L. for European Space Agency (ESA), Camino bajo del Castillo, s/n, Urbanizacion Villafranca del Castillo, Villanueva de la Ca\~nada, 28692 Madrid, Spain\label{aff70}
\and
Institut de F\'{i}sica d'Altes Energies (IFAE), The Barcelona Institute of Science and Technology, Campus UAB, 08193 Bellaterra (Barcelona), Spain\label{aff71}
\and
European Space Agency/ESTEC, Keplerlaan 1, 2201 AZ Noordwijk, The Netherlands\label{aff72}
\and
DARK, Niels Bohr Institute, University of Copenhagen, Jagtvej 155, 2200 Copenhagen, Denmark\label{aff73}
\and
Waterloo Centre for Astrophysics, University of Waterloo, Waterloo, Ontario N2L 3G1, Canada\label{aff74}
\and
Department of Physics and Astronomy, University of Waterloo, Waterloo, Ontario N2L 3G1, Canada\label{aff75}
\and
Perimeter Institute for Theoretical Physics, Waterloo, Ontario N2L 2Y5, Canada\label{aff76}
\and
Space Science Data Center, Italian Space Agency, via del Politecnico snc, 00133 Roma, Italy\label{aff77}
\and
Centre National d'Etudes Spatiales -- Centre spatial de Toulouse, 18 avenue Edouard Belin, 31401 Toulouse Cedex 9, France\label{aff78}
\and
Institute of Space Science, Str. Atomistilor, nr. 409 M\u{a}gurele, Ilfov, 077125, Romania\label{aff79}
\and
Dipartimento di Fisica e Astronomia "G. Galilei", Universit\`a di Padova, Via Marzolo 8, 35131 Padova, Italy\label{aff80}
\and
INFN-Padova, Via Marzolo 8, 35131 Padova, Italy\label{aff81}
\and
Instituto de F\'isica Te\'orica UAM-CSIC, Campus de Cantoblanco, 28049 Madrid, Spain\label{aff82}
\and
Institut de Recherche en Astrophysique et Plan\'etologie (IRAP), Universit\'e de Toulouse, CNRS, UPS, CNES, 14 Av. Edouard Belin, 31400 Toulouse, France\label{aff83}
\and
Universit\'e St Joseph; Faculty of Sciences, Beirut, Lebanon\label{aff84}
\and
Departamento de F\'isica, FCFM, Universidad de Chile, Blanco Encalada 2008, Santiago, Chile\label{aff85}
\and
Department of Physics and Helsinki Institute of Physics, Gustaf H\"allstr\"omin katu 2, University of Helsinki, 00014 Helsinki, Finland\label{aff86}
\and
Departamento de F\'isica, Faculdade de Ci\^encias, Universidade de Lisboa, Edif\'icio C8, Campo Grande, PT1749-016 Lisboa, Portugal\label{aff87}
\and
Instituto de Astrof\'isica e Ci\^encias do Espa\c{c}o, Faculdade de Ci\^encias, Universidade de Lisboa, Tapada da Ajuda, 1349-018 Lisboa, Portugal\label{aff88}
\and
Cosmic Dawn Center (DAWN)\label{aff89}
\and
Niels Bohr Institute, University of Copenhagen, Jagtvej 128, 2200 Copenhagen, Denmark\label{aff90}
\and
Universidad Polit\'ecnica de Cartagena, Departamento de Electr\'onica y Tecnolog\'ia de Computadoras,  Plaza del Hospital 1, 30202 Cartagena, Spain\label{aff91}
\and
European University of Technology EUt+, European Union\label{aff92}
\and
Institute of Space Sciences (ICE, CSIC), Campus UAB, Carrer de Can Magrans, s/n, 08193 Barcelona, Spain\label{aff93}
\and
Institut d'Estudis Espacials de Catalunya (IEEC),  Edifici RDIT, Campus UPC, 08860 Castelldefels, Barcelona, Spain\label{aff94}
\and
Caltech/IPAC, 1200 E. California Blvd., Pasadena, CA 91125, USA\label{aff95}
\and
Astronomisches Rechen-Institut, Zentrum f\"ur Astronomie der Universit\"at Heidelberg, M\"onchhofstr. 12-14, 69120 Heidelberg, Germany\label{aff96}
\and
Institut d'Astrophysique de Paris, 98bis Boulevard Arago, 75014, Paris, France\label{aff97}
\and
Department of Physics and Astronomy, University of British Columbia, Vancouver, BC V6T 1Z1, Canada\label{aff98}}    
\date{Received XXX; accepted YYY}

\begin{document}
\label{firstpage}

\abstract
{Dwarf galaxies in dense cluster environments are susceptible to tidal interactions that can alter their morphology and kinematics. Boxy isophotes are well studied in massive galaxies but remain poorly understood in dwarfs.}
{We aim to identify and characterise boxy/peanutty dwarf galaxies in the Perseus cluster and determine the origin of their isophotal shapes.}
{Using \textit{Euclid} Early Release Observations of the Perseus cluster, we present a cumulative light fraction method for robustly measuring the isophotal shape parameter $c_4$, particularly suited to low surface brightness regimes. From the $\sim1100$ cataloged dwarfs, we select a clean sample of $\sim190$ early-type systems sufficiently resolved and free of contamination for reliable $c_4$ measurement. Observed trends are interpreted through comparison with mock \textit{Euclid} observations of gravitational $N$-body simulations of tidally transformed dwarfs.}
{From this clean sample, we identify 13 dwarfs with significantly boxy isophotes ($c_4 < -0.0175$), confirmed through visual inspection. These galaxies lack visible thin disks, lie on the cluster red sequence, and show no preferential spatial concentration within Perseus (Kolmogorov--Smirnov $p = 0.65$ versus the wider dE sample). We find a significant anticorrelation between $c_4$ and effective radius: larger galaxies exhibit more boxy (more negative $c_4$) isophotes. We caution that the linear (Pearson) coefficient is sensitive to the two most boxy galaxies; the rank-based (Spearman) coefficient remains significant ($r_s = -0.59$, $p = 0.035$). The parameter $c_4$ shows no correlation with S\'ersic index or concentration. An analogous size--shape anticorrelation is recovered in the simulations, where inner regions are dominated by box orbits associated with a triaxial peanut structure and outer regions by short-axis tube orbits.}
{Boxy/peanutty dwarf galaxies in Perseus are tidally transformed remnants of moderately rotating progenitors, with boxy isophotes tracing inner box-orbit-dominated peanut structures. The size--shape correlation arises from viewing geometry: face-on orientations reveal the rectangular profile of the elongated triaxial structure (large and boxy), while edge-on views yield rounder, compact morphologies. Our sample thus represents an orientation-selected subset of tidally transformed, peanutty dwarfs in the cluster.}



\maketitle
\nolinenumbers  
\section{Introduction}
\label{sect:intro}

Dwarf galaxies ($V$-band absolute magnitude $M_V \gtrsim -18$) constitute the most numerous class of galaxies in the Universe and serve as critical probes of galaxy formation and evolution in hierarchical structure formation models \citep{2001ApSSS.277..231G, 2012AJ....144....4M}. In cluster environments, their low masses make them particularly sensitive to external perturbations, including ram pressure stripping \citep{1972ApJ...176....1G} and tidal interactions \citep{2001ApJ...559..754M}, most notably tidal harassment (the cumulative effect of multiple high-speed gravitational encounters within a dense cluster; \citealt{1996Natur.379..613M}). Understanding how these environmental processes imprint observable signatures on dwarf galaxy morphologies provides key insights into the physical mechanisms that regulate galaxy evolution in dense environments.

Among the morphological properties that have proven valuable for diagnosing formation and evolutionary pathways, isophotal shapes stand out as particularly informative. The shapes of galaxy isophotes, deviations of galaxy contours from perfect ellipses, can be quantified through Fourier analysis of deviations from fitted ellipses. The traditional parametrisation, introduced by \citet{1988A&AS...74..385B}, expresses these deviations as normalised Fourier coefficients $a_4/a$. Negative values indicate ``boxy'' isophotes and positive values indicate ``disky'' isophotes \citep{1989A&A...217...35B}. In this work, we use the analogous coefficient $c_4$, measured via a cumulative light fraction technique (see Sect.~\ref{sect:isophote_impl} for the precise definition). In massive elliptical galaxies, deviations from elliptical isophotes have been linked to the underlying orbital structure of stars and the distribution of dark matter \citep{1989ARA&A..27..235K}. Boxy isophotes are commonly observed in pressure-supported remnants potentially shaped by mergers, while disky isophotes are associated with rotationally supported systems. However, the detailed physical mechanisms connecting isophotal shapes to specific formation histories are not fully resolved \citep{2005MNRAS.359..949B, 2006MNRAS.372..839N, 2009ApJS..182..216K}.

While isophotal shape analysis has a rich history in the context of massive galaxies \citep{1989A&A...217...35B, 1989ARA&A..27..235K, 2006MNRAS.370.1339H, 2009ApJS..182..216K}, its application to dwarf galaxies has been more limited. One of the earliest systematic studies of dwarf galaxy isophotes was conducted by \citet{1999ApJ...517..650R} for early-type dwarfs in the Virgo cluster, concluding that the dwarf elliptical galaxies in Virgo may have the
same dichotomy between rotation-supported systems and systems supported by velocity anisotropy that is seen in the more luminous systems. In Perseus specifically, \citet{2003AJ....125...66C} measured $a_4$ for the brightest 14 low-mass cluster galaxies in the cluster core using ground-based WIYN imaging, reporting small deviations broadly consistent with elliptical shapes; the depth and resolution of \textit{Euclid} VIS now permit an analogous analysis on a sample more than an order of magnitude larger and extending to lower surface brightness. Subsequent studies \citep{2003A&A...407..121B, 2014ApJ...787..102C} extended these measurements to fainter surface brightness regimes and confirmed significant radial variations in isophotal properties within individual galaxies. More recently, large surveys such as the Fornax Deep Survey \citep{2018A&A...620A.165V, 2019A&A...625A.143V} and the MATLAS survey \citep{Poulain2021} have explored dwarf morphologies across different environments, revealing environmental trends in structural parameters and nucleation fractions. Despite this progress, the physical interpretation of boxy isophotes in dwarfs remains uncertain, with competing scenarios including dry merger remnants \citep{2006MNRAS.372..839N} and tidally transformed systems \citep[hereafter S21]{2021ApJ...912..149S}.

A key challenge in measuring isophotal shapes for dwarf galaxies is their intrinsically low surface brightness, which renders traditional fixed surface brightness isophote methods (approaches that measure isophotal shapes by tracing galaxy contours at a fixed surface brightness threshold) unreliable or susceptible to noise contamination. Furthermore, connecting observed morphologies to formation mechanisms requires not only robust measurements but also careful comparison with theoretical predictions. Synthetic observations of simulated galaxies provide a crucial tool for disentangling projection effects from intrinsic structural variations and for testing whether observed morphological trends are consistent with specific evolutionary scenarios.

Addressing these challenges requires both deep, high-resolution imaging and a well-characterised dwarf population. The \textit{Euclid} space telescope \citep{2025A&A...697A...1E}, with its Visual Instrument (VIS) delivering diffraction-limited imaging at $0\farcs1$ pixel$^{-1}$ and exceptional surface brightness sensitivity, provides an ideal instrument for systematic morphological studies of dwarfs (described fully in Sect.~\ref{sect:data}). The Perseus cluster was one of the targets selected for the \textit{Euclid} Early Release Observations (ERO) programme \citep{EROcite, EROData}. Perseus itself, with a velocity dispersion of $\sigma_{\rm cl} \approx 1300$ km s$^{-1}$ \citep{1983AJ.....88..697K} and a virial mass comparable to Virgo, is a typical rich cluster environment where environmental processing operates efficiently. The ERO catalog compiled by \citet{EROPerseusDGs}, supported by the overview analysis of \citet{EROPerseusOverview}, contains approximately 1100 dwarf galaxies, making Perseus only the second cluster (after Virgo) to receive systematic morphological analysis of its dwarf population at this level of detail and depth.

In this paper, we exploit the \textit{Euclid} ERO data to conduct a systematic analysis of isophotal shapes in Perseus cluster dwarf galaxies. We develop a cumulative light fraction approach to measure the fourth-order Fourier coefficient $c_4$, specifically designed for low surface brightness systems, and interpret the observed morphological trends through comparison with gravitational $N$-body simulations of tidally transformed dwarfs that develop peanut\footnote{Although the simulation literature sometimes refers to such elongated inner structures as ``bars,'' we deliberately avoid this term: bars are sub-structural components of disk galaxies driven by internal rotational instabilities, whereas the structures studied here are externally induced by tidal forcing and constitute the entire stellar body of the dwarf. We use ``peanut structure'' or ``elongated triaxial structure'' throughout to reflect their tidal origin.} structures.

The paper is organised as follows. Section~\ref{sect:data} describes the \textit{Euclid} ERO data and our Perseus dwarf galaxy sample. Section~\ref{sect:methodology} details our morphological measurement pipeline, including the cumulative light fraction technique for $c_4$ measurements. Section~\ref{sect:results} presents the identification and structural properties of our boxy dwarf sample and examines correlations between boxiness and other galaxy parameters. Section~\ref{sect:interpretation} interprets these findings through comparison with simulations, examining the role of projection effects, orbital structure, and evolutionary timescales, and evaluates alternative formation channels. We summarise our conclusions in Sect.~\ref{sect:summary}.
\section{\textit{Euclid} observations of the Perseus cluster}
\label{sect:data}

\subsection{ERO observations of Perseus}

This study utilises data from the ERO programme, specifically targeting the Perseus cluster field. \Euclid{} is equipped with two instruments: the Visual Instrument \citep{2025A&A...697A...2E} covering wavelengths of 550--900 nm, and the Near Infrared Spectrometer and Photometer (NISP; \citealt{2025A&A...697A...3E}) providing imaging in the \YE, \JE, and \HE bands. The VIS instrument delivers imaging with a pixel scale of $0\farcs1$ per pixel, while NISP has a pixel scale of $0\farcs3$. The post-resampling \IE mosaics from the dedicated ERO reduction \citep{EROData} used here have a measured PSF FWHM of $0\farcs158$ (as determined via \texttt{PSFEx}; the unresampled native value is $0\farcs136$, i.e.\ near the diffraction limit). The NISP PSF FWHMs are $0\farcs48$, $0\farcs49$, and $0\farcs50$ in the \YE, \JE, and \HE bands, respectively. These capabilities make \textit{Euclid} particularly well-suited for detailed morphological studies of galaxies, even at the distances of nearby clusters like Perseus.

The Perseus cluster ERO data consist of deep observations covering approximately 0.7 square degrees centred on the cluster core \citep{EROPerseusOverview}. The observations consisted of four reference observing sequences (ROSs), each containing four dithered exposures of 566 seconds in the \IE filter and four exposures of 87.2 seconds each in the NISP \YE, \JE, and \HE filters \citep{2022A&A...662A.112E}. The data were reduced using the dedicated pipeline specifically developed for the ERO programme \citep{EROData}, which includes bias subtraction, flat fielding, cosmic ray removal, astrometric calibration (achieving 8 mas precision for VIS and 15 mas for NISP using Gaia DR3 as reference), and photometric calibration (with absolute zero point accuracy at the percent level). The final reduced \IE images reach a 5$\sigma$ point source depth of 28.0 mag and surface brightness limits for radially integrated galaxy profiles of 30.1 mag arcsec$^{-2}$, enabling the detection and detailed analysis of dwarf galaxies at the distance of Perseus. While the morphological analysis in this work primarily relies on the higher-resolution \IE data, the NISP observations provide valuable colour information for studying stellar populations and confirming cluster membership.

\subsection{Dwarf galaxy sample}
Our sample of dwarf galaxies is drawn from the catalog of Perseus cluster members compiled in \citet{EROPerseusDGs}, which was based on visual inspection of the multi-wavelength ERO data. This catalog was constructed through visual classification by seven independent classifiers who examined each candidate galaxy for morphological features, nucleation, globular cluster content, and signs of tidal disturbance. Of the resulting candidates, 264 were flagged as dwarfs by all seven classifiers and 859 by at least four; the final $\sim 1100$ figure corresponds to a $\geq 5/7$ agreement cut.

The complete catalog contains approximately 1100 dwarf galaxy candidates within the ERO field of view. Of these, 1061 (96\%) are classified as dwarf ellipticals (dE), while 39 (4\%) are classified as dwarf irregulars (dI). Additionally, 581 (53\%) show evidence of nucleation, 282 (26\%) are classified as globular cluster-rich, and 64 (6\%) exhibit disturbed morphologies indicative of tidal interactions. 

The sample includes 93 ultra-diffuse galaxies (UDGs, 8.5\% of the total dwarf population), defined following \citet{2015ApJ...798L..45V} as galaxies with central surface brightness $\mu_{g,0} \geq 24$ mag arcsec$^{-2}$ and $R_{\rm e} \geq 1.5$ kpc. This UDG fraction is higher than found in other nearby clusters such as Coma (5.4\%, \citealt{2024ApJS..271...52Z}) or the MATLAS survey (2.7\%, \citealt{2021A&A...654A.105M}), likely due to the exceptional surface brightness depth of the ERO data and the high angular resolution that enables clear distinction between faint dwarfs and background sources.

For comparison with previous work, \citet{EROPerseusDGs} note that their dwarf sample consists of galaxies with $M_V > -18$ at the adopted distance of 72 Mpc for the Perseus cluster. The catalog completeness begins to decline rapidly at $M_{\IE} = -12$, dropping to 50\% at $M_{\IE} = -11$ \citep{EROPerseusOverview}.

For this morphological study, we analyse the subset of these dwarf galaxies that have successful measurements from our pipeline, as described in Sect.~\ref{sect:methodology}.

\section{A robust \texorpdfstring{$c_4$}{c4} measurement technique for low surface brightness systems}
\label{sect:methodology}

Our pipeline integrates \texttt{MTObjects} \citep{Teeninga2015} and the Python morphology code \texttt{statmorph} \citep{Rodriguez-Gomez2019} to analyse dwarf galaxy morphology in \textit{Euclid} data. We have modified the latter to include custom boxiness/diskiness measurements.
Here, we detail the full workflow as implemented in our code.

\begin{figure*}
    \centering
    \includegraphics[width=0.7\textwidth]{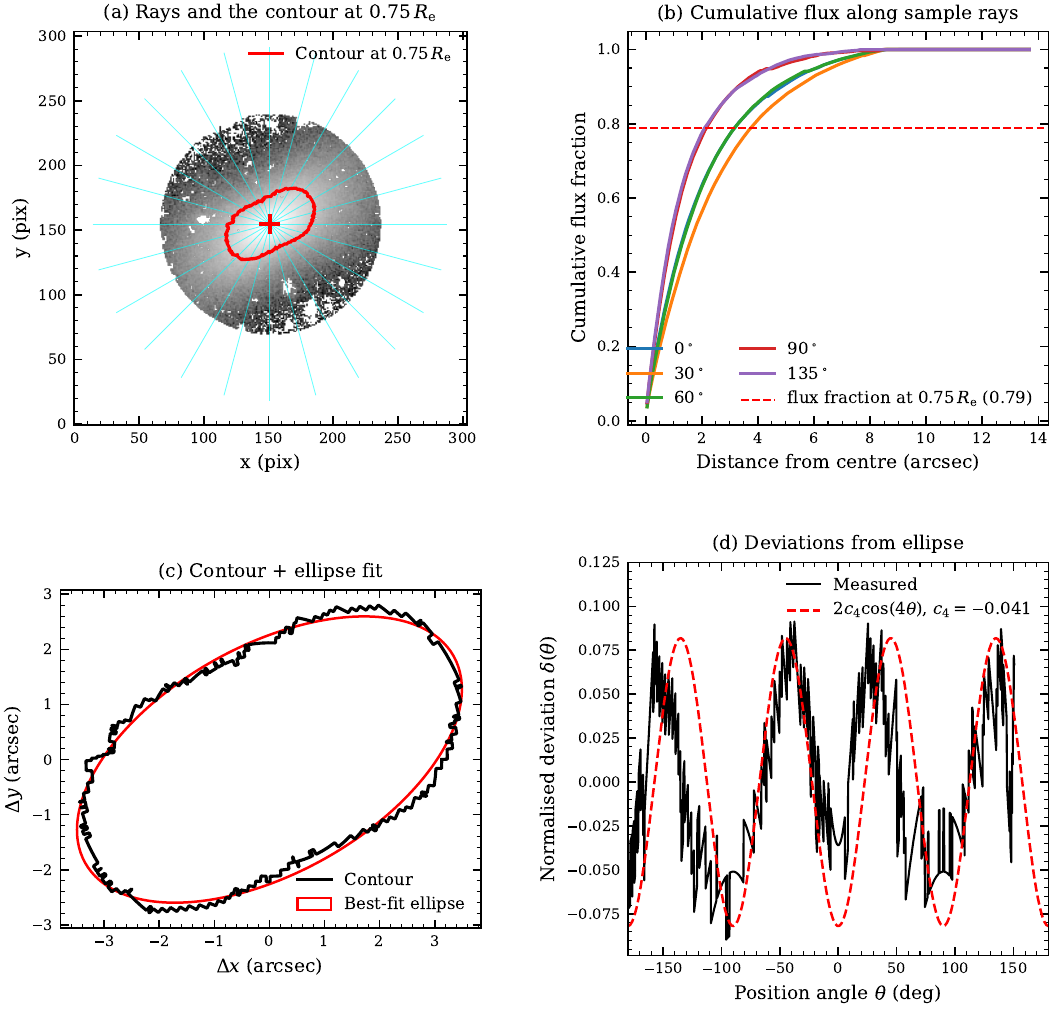}
    \caption{Illustration of the $c_4$ measurement on a representative boxy dwarf (EDwC-0515, $c_4 = -0.041$ at $0.75\,R_{\rm e}$). Panel (a): \IE-band stamp with radial rays (cyan) emanating from the determined centre (red cross) and the contour (red) at $0.75\,R_{\rm e}$. Panel (b): cumulative-flux profiles along five sample rays at the indicated position angles; the dashed red line marks the flux fraction whose interpolated radius lies at $0.75\,R_{\rm e}$. Panel (c): the extracted contour (black) and corresponding best-fit ellipse (red). Panel (d): the normalised radial deviations $\delta(\theta)$ from the ellipse (black), with the $2 c_4 \cos(4\theta)$ component (red dashed) shown for comparison.}
    \label{fig:c4_method}
\end{figure*}

\subsection{Pipeline overview}

The complete analysis pipeline processes the galaxy cutouts through multiple stages. First, we load each cutout and perform a quality assessment to ensure adequate data coverage (Sect.~\ref{sect:cutouts}). Next, we apply the \texttt{MTObjects} algorithm \citep{Teeninga2015} for source detection and segmentation (Sect.~\ref{sect:mtobjects}). We then identify which detected segment corresponds to the target dwarf galaxy using a composite scoring system (Sect.~\ref{sect:identification}). Following this identification, we determine the galaxy centre and refine the segmentation map to ensure reliable measurements (Sect.~\ref{sect:center}). Finally, we perform the morphological analysis using \texttt{statmorph} with our custom boxiness/diskiness measurements (Sect.~\ref{sect:statmorph}). This integrated approach ensures consistent processing across our sample of Perseus cluster dwarf galaxies and enables reliable comparison between different objects.

\subsection{Cutout preparation}
\label{sect:cutouts}

For each dwarf galaxy in the ERO catalog, we extract cutout images from the \textit{Euclid} \IE and NISP mosaics. The size of each cutout is determined adaptively based on the effective radius of the target galaxy, ensuring coverage extends to $9\,R_{\rm e}$ \citep{Poulain2021}. A quality assessment is performed on each cutout, rejecting those with more than 20\% missing data (NaN values) to ensure reliable morphological measurements. The pipeline supports processing in all \textit{Euclid} bands (\IE, \YE, \JE, \HE), with \IE data used for the primary morphological analysis due to its superior spatial resolution (see Sect.~\ref{sect:data}).

\subsection{Source detection with \texttt{MTObjects}}
\label{sect:mtobjects}

Although the target galaxy positions come from the \citet{EROPerseusDGs} catalog, \texttt{MTObjects} serves a distinct purpose here: rather than finding sources, it provides a pixel-level segmentation map that delineates the galaxy footprint and separates it from neighbouring objects and background sources. We do not use the standard MER segmentation maps because \texttt{MTObjects} has been shown to outperform standard source-extraction tools at recovering the faint outskirts of low surface brightness galaxies \citep{Haigh2021}. \texttt{MTObjects} constructs a max-tree representation of the image and applies statistical significance testing to identify structures against the background noise. This approach performs well in the low surface brightness regime characteristic of dwarf galaxies, as demonstrated by benchmarks against several common detection codes \citep{Haigh2021}. Our implementation uses \texttt{move\_factor}$= 0.3$ to balance detection of extended features while separating neighboring sources.

\subsection{Main galaxy identification}
\label{sect:identification}

Although each cutout is centred on the catalog position, \texttt{MTObjects} typically returns multiple segments per cutout, so the target is not guaranteed to be the largest or most central. We pre-filter candidates to match the expected dwarf regime (mean \IE surface brightness $\ge 22.0$~mag~arcsec$^{-2}$, \IE magnitude $15.5$--$25.0$, equivalent radius $>15$~pixels, axis ratio $<10$) and rank the survivors via a composite
\begin{align}
\text{Score} &= 0.6 \, \text{Centrality} + 0.2 \, \text{Area Score} + 0.2 \, \text{Edge Factor},
\end{align}
where Centrality, Area Score, and Edge Factor are normalised metrics for proximity to cutout centre, segment size, and isolation from image edges, respectively. A worked example of this scoring for a representative galaxy (EDwC-0011) is given in Appendix~\ref{app:mtobjects_example}.

\subsection{Centre determination and segmentation map refinement}
\label{sect:center}

We employ a hill-climbing algorithm inspired by \texttt{AutoProf}, an automated non-parametric light-profile pipeline \citep{2021MNRAS.508.1870S}, for centre determination. Accurate centring is critical, as the $c_4$ coefficient is sensitive to centre choice (a quantitative assessment is given in Appendix~\ref{app:centre_sensitivity}). Bright compact sources (globular clusters, foreground stars) are naturally excluded via \texttt{MTObjects} segmentation; additional pixel clipping removes residual outliers that could bias convergence.

The segmentation map is then restricted to within $2.5\,R_{\rm e}$ of the determined centre, to limit the analysis to the dwarf's main body and exclude any contaminating segments at large radii. To identify internal gaps in the segmentation map (caused by masked foreground or background sources within the dwarf footprint), we generate a hole-filled version of the segment using \texttt{scipy.ndimage.binary\_fill\_holes} and take its difference with the original map; the resulting hole regions are flagged if their equivalent radius exceeds $0.2\,R_{\rm e}$.

\subsection{Morphological analysis with \texttt{statmorph}}
\label{sect:statmorph}

We employ \texttt{statmorph} to compute non-parametric diagnostics (concentration, asymmetry, smoothness, Gini, $M_{20}$) and fit Sérsic profiles, using primarily \IE data for its superior resolution. A central contribution of this work is the development of a robust method to measure and analyse isophotal shapes in low surface brightness dwarf galaxies, implemented as an extension to the \texttt{statmorph} package (Sect.~\ref{sect:isophote_impl}).

\subsubsection{Implementation of isophotal shape analysis}
\label{sect:isophote_impl}

\begin{figure*}
    \centering
    \includegraphics[width=0.8\textwidth]{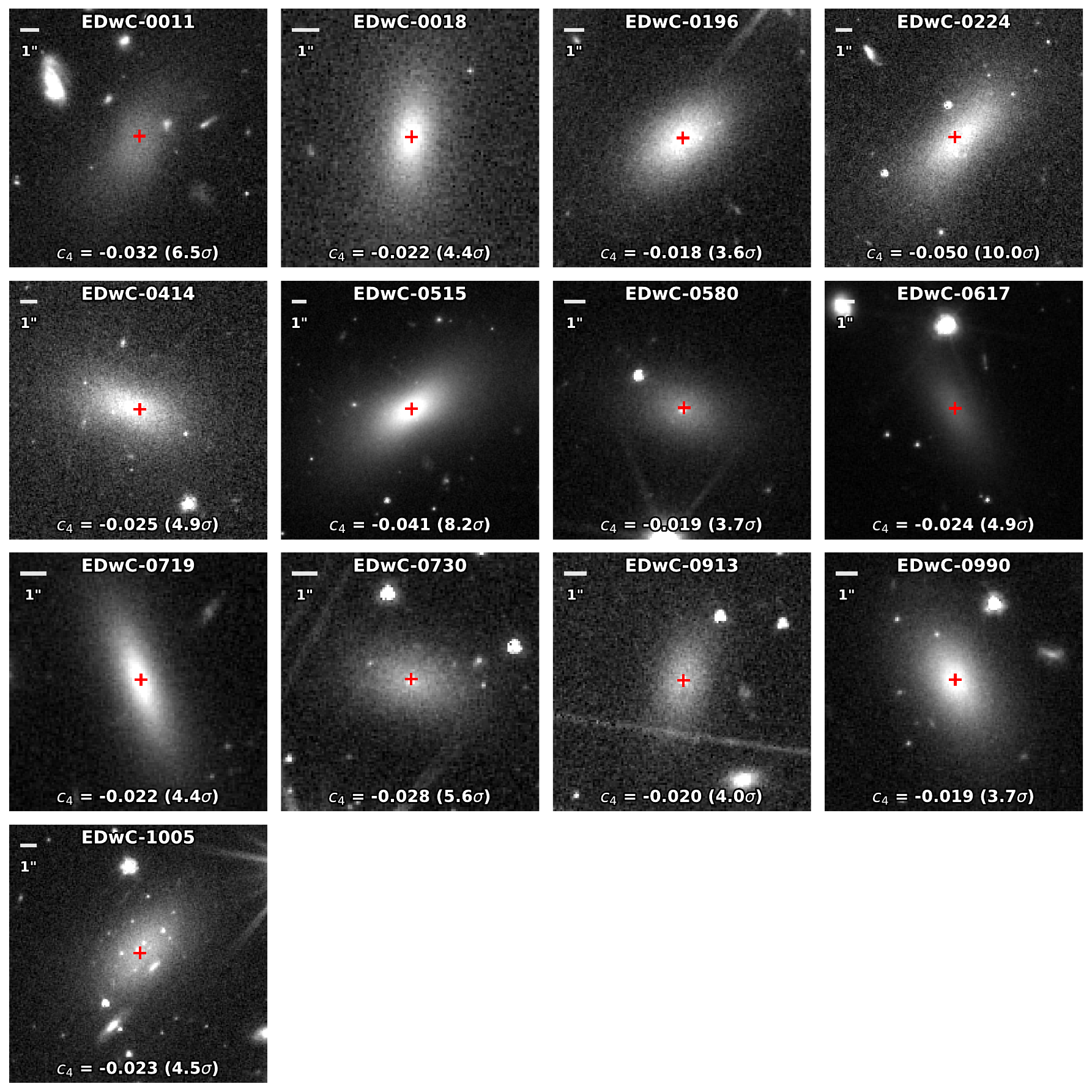}
    \caption{\IE-band images of the 13 Perseus dwarfs with significantly boxy isophotes ($c_4 < -0.0175$), shown with a logarithmic intensity scaling between the 0.5 and 99.5 percentile of positive pixel values per cutout. Each panel lists the measured $c_4$ and the corresponding detection significance $|c_4|/\sigma_{\rm typ}$, adopting the median validation $\sigma_{\rm typ} = 0.005$ from Appendix~\ref{app:validation}; pipeline-determined centres are marked with red crosses. None of the 13 dwarfs exhibit visible disk features or spiral structure.}
    \label{fig:boxy_montage}
\end{figure*}

Galaxy isophotal shapes are traditionally characterised by fitting ellipses and analysing deviations from perfect elliptical forms using Fourier decomposition. These deviations can be expressed as
\begin{equation}
r(\theta) = r_0 \left[1 + \sum_{n=1}^{\infty} a_n \cos(n\theta) + b_n \sin(n\theta) \right] \,,
\end{equation}
where $r(\theta)$ represents the radius as a function of azimuthal angle $\theta$, $r_0$ is the average isophotal radius, and $a_n$ and $b_n$ are the Fourier coefficients. Our implementation extends this traditional approach with a multi-radial measurement strategy designed for low surface brightness dwarf galaxies. Figure~\ref{fig:c4_method} illustrates the workflow on a representative boxy dwarf: panel~(a) shows the radial rays and the extracted contour at $0.75\,R_{\rm e}$, panel~(b) the cumulative-flux profiles along five sample rays, panel~(c) the contour and best-fit ellipse, and panel~(d) the residual deviations $\delta(\theta)$ with the $2c_4\cos(4\theta)$ component overlaid.

\paragraph{Contour extraction using cumulative light fractions.}
Rather than using fixed surface brightness levels, we extract contours at cumulative light fractions. For each galaxy, we compute contours at 20 equally-spaced flux levels ranging from 0.2 to 0.9 of the total integrated flux. This flux-based approach is particularly advantageous when analysing a heterogeneous sample spanning a wide range of intrinsic surface brightness. Fixed surface brightness thresholds must be tuned to each galaxy's brightness: a threshold appropriate for a bright dwarf may fall in the noisy outskirts or below the detection limit for a fainter one. Cumulative light fractions define contour levels relative to each galaxy's own flux distribution, so the same set of fractional levels (0.2--0.9) automatically samples the detectable footprint of every galaxy in the sample without per-object tuning. These fractions are ordered from the inner to the outer parts of the galaxy, and span the bulk of each galaxy's light while excluding the innermost contours, where PSF smoothing affects the isophotal shape, and the outermost contours, where the cumulative profile flattens.

The contour extraction uses radial ray profiles emanating from the galaxy centre. We generate rays spanning the full azimuthal range, with the number of rays determined adaptively based on image size: we compute the maximum possible radius from centre to image boundary, calculate the circumference (in pixels) at this radius, and use twice that number of rays so that the angular sampling is twice as fine as the pixel grid at the largest radius. For each ray, we compute the cumulative flux profile and determine the radius at which it crosses the target flux fraction through linear interpolation.

To ensure robust measurements, we implement ray filtering to reject rays crossing holes in the segmentation map or encountering masked regions. A ray is excluded if it contains zero-valued or NaN pixels between its first and last nonzero values, as such discontinuities would bias the cumulative flux calculation. This filtering is critical for dwarf galaxies where foreground stars, background galaxies, or diffraction spikes may create localized masking artifacts.

\begin{figure*}
    \centering
    \includegraphics[width=0.8\textwidth]{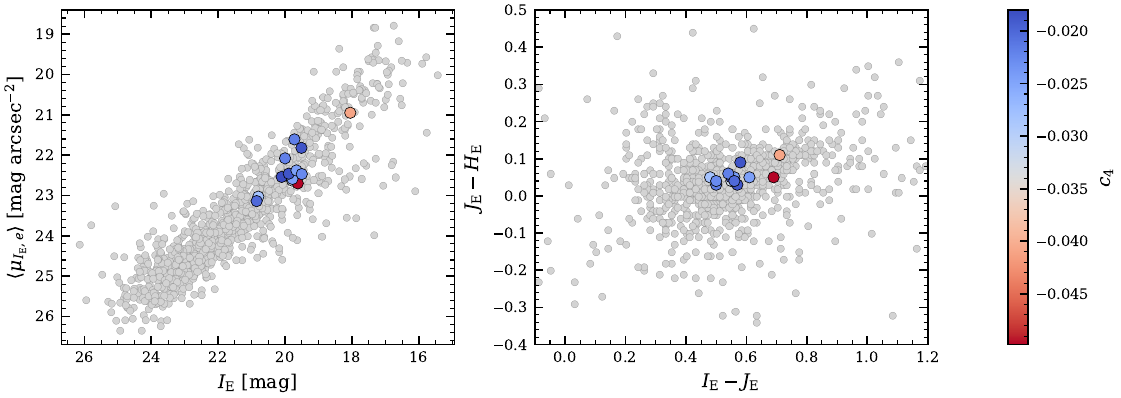}
    \caption{Photometric context of the 13 boxy dwarf candidates (coloured markers; $c_4 < -0.0175$, with more negative $c_4$ indicating stronger boxiness) superimposed on the parent population of 1061 Perseus early-type dwarfs from the \citet{EROPerseusDGs} catalog (grey points; 1033 in the right panel, restricted to galaxies with all three \textit{Euclid} bands measured). \emph{Left}: mean effective surface brightness $\langle \mu_{\IE,\rm e} \rangle$ versus \IE magnitude. \emph{Right}: near-infrared colour-colour diagram showing $\JE-\HE$ versus $\IE-\JE$.}
    \label{fig:vis_nir_context}
\end{figure*}

\paragraph{Ellipse fitting and Fourier decomposition.}
For each extracted contour, we fit an ellipse with its centre fixed at the galaxy centre. The fitting procedure uses a polar representation and determines the semi-major axis $a$, semi-minor axis $b$, and position angle $\phi$ that best match the contour points through least-squares minimisation using \texttt{scipy.optimize.curve\_fit}. We then compute normalised radial deviations from the best-fit ellipse:
\begin{equation}
\delta(\theta) = \frac{r_{\rm obs}(\theta) - r_{\rm ell}(\theta)}{r_{\rm eq}} \,,
\end{equation}
where $r_{\rm obs}(\theta)$ is the observed radius at position angle $\theta$, $r_{\rm ell}(\theta)$ is the radius predicted by the fitted ellipse, and $r_{\rm eq} = \sqrt{ab}$ is the equivalent radius (geometric mean of the semi-axes). This normalization ensures deviations are scale-independent and comparable across galaxies of different sizes and ellipticities.

The $c_4$ coefficient is extracted via cross-correlation with the fourth harmonic:
\begin{equation}
c_4 = \frac{2}{N} \sum_{i=1}^{N} \delta(\theta_i) \cos(4\theta_i) \,,
\end{equation}
where $N$ is the number of contour points and $\theta_i$ are measured relative to the ellipse major axis. This formulation extracts the amplitude of the $\cos(4\theta)$ component, with $c_4 < 0$ indicating boxy isophotes and $c_4 > 0$ indicating disky isophotes. Our $c_4$ follows the definition of \citet{1989AJ.....98..538F}, measuring the relative radial deviation from the best-fit ellipse normalised by the equivalent radius $r = \sqrt{ab}$, but applied to cumulative light fraction contours rather than fixed surface brightness isophotes. It relates to the $a_4/a$ of \citet{1988A&AS...74..385B} as $a_4/a = c_4\,\sqrt{b/a}$ \citep{1999BaltA...8..535M}.

\paragraph{Output measurements.}
Our pipeline calculates $c_4$ at each of the 20 cumulative light fractions, producing a radial profile. To obtain a single characteristic value for each galaxy, we interpolate this profile and measure $c_4$ at $0.75\,R_{\rm e}$. This radius is far enough from the centre that PSF smoothing does not dominate the isophotal shape, yet well within the high signal-to-noise interior of the galaxy. This characteristic $c_4$ value is provided alongside concentration index, asymmetry, smoothness, Gini coefficient, $M_{20}$, Sérsic index, and ellipticity, enabling systematic comparison of isophotal shapes with other morphological diagnostics across the Perseus dwarf galaxy sample.

Our cumulative light fraction approach is also less sensitive to localised features or measurement artefacts than fixed surface brightness methods. By applying it to \textit{Euclid} observations and selecting galaxies with boxy morphologies, we can efficiently assemble a statistically significant sample of these rare objects.

All images used in this analysis are background-subtracted by the ERO data reduction pipeline \citep{EROData}. We validated our method using synthetic galaxies with \textit{Euclid}-realistic noise properties matched to the observed sample (see Appendix~\ref{app:validation} for details). The validation confirms a tight, linear relationship between input isophotal shape and measured $c_4$. Crucially, for galaxies with boxy isophotes in the range of our sample ($c_4 \sim -0.02$ to $-0.05$), the per-galaxy measurement uncertainty is substantially smaller than the measured signal, confirming that our detections are robust.
\section{The boxy dwarf population in Perseus}
\label{sect:results}

\subsection{Identification and structural properties}
We begin by constructing a clean sample of early-type dwarfs for which robust $c_4$ measurements can be obtained, and then identify the boxy subset within it. Starting from $\sim1100$ galaxies in the \citet{EROPerseusDGs} catalog, 1061 are classified as early-type (dE) systems. We filter these for adequate angular size ($R_{\rm e} > 1\farcs5$, i.e.\ 15 pixels), surface brightness ($\mu < 24.75$~mag~arcsec$^{-2}$), photometric consistency ($|\Delta m| < 0.5$ between \texttt{MTObjects} and catalog photometry), and clean segmentation maps (no large holes from masked contaminants). This yields a clean sample of 183 early-type dwarfs with reliable $c_4$ measurements. Of these, 13 have significantly boxy isophotes ($c_4 < -0.0175$), confirmed through visual inspection, and are presented in Fig.~\ref{fig:boxy_montage}.

The threshold $c_4 < -0.0175$ was determined empirically: galaxies below this value are visually and unambiguously boxy upon direct image examination, and all 13 were independently confirmed through visual inspection. The clean sample $c_4$ distribution (Appendix~\ref{app:c4_distribution}) is approximately Gaussian with $\sigma = 0.013$, broader than the per-galaxy measurement uncertainty of $0.003$--$0.008$ (Appendix~\ref{app:validation}), confirming that the observed scatter reflects genuine variation in isophotal shape. Each boxy dwarf is individually a $3$--$15\sigma$ detection.

The structural and photometric properties of the 13 galaxies are tabulated in Table~\ref{Tab:sample}, and Fig.~\ref{fig:vis_nir_context} shows them in the context of the wider Perseus dwarf sample. In brief, the galaxies span $R_{\rm e} \approx 0.86$--$2.07$~kpc, absolute magnitudes $M_{\IE} \approx -13.9$ to $-16.7$, and boxiness values $c_4 \approx -0.018$ to $-0.050$, forming a photometrically homogeneous population on the cluster red sequence. The boxy dwarfs show no preferential spatial concentration within the cluster, their projected clustercentric distances (median $15\farcm1$) are statistically indistinguishable from the full early-type dwarf sample (median $17\farcm9$; two-sample Kolmogorov--Smirnov test $p = 0.65$).

\begin{table}[ht]
    \caption{Structural and photometric properties for the 13 boxy dwarf galaxy candidates ($c_4 < -0.0175$).}
    \centering
    \small
    \begin{tabular}{lcc}
    \hline
    \noalign{\smallskip}
    Parameter & Mean $\pm$ Std & Range \\
    \noalign{\smallskip}
    \hline
    \noalign{\smallskip}
    $R_{\rm e}$ (arcsec) & $3.93 \pm 1.20$ & $[2.52, 6.09]$ \\
    $R_{\rm e}$ (kpc) & $1.33 \pm 0.41$ & $[0.86, 2.07]$ \\
    $c_4$ ($0.75\,R_{\rm e}$) & $-0.026 \pm 0.010$ & $[-0.050, -0.018]$ \\
    \noalign{\smallskip}
    $\IE$ & $19.79 \pm 0.68$ & $[18.05, 20.85]$ \\
    $M_{\IE}$ & $-14.93 \pm 0.67$ & $[-16.71, -13.90]$ \\
    $\langle \mu_{\IE,\rm e} \rangle^a$ & $22.34 \pm 0.60$ & $[20.95, 23.14]$ \\
    $\IE-\JE$ & $0.57 \pm 0.07$ & $[0.48, 0.71]$ \\
    \noalign{\smallskip}
    $b/a$ & $0.54 \pm 0.16$ & $[0.30, 0.95]$ \\
    \noalign{\smallskip}
    \hline
    \end{tabular}
    \label{Tab:sample}
    \tablefoot{Physical sizes assume $d = 72$ Mpc. $^a$In mag arcsec$^{-2}$.}
\end{table}

\subsection{The size--shape correlation}
\label{sect:correlations}

\begin{figure}
    \centering
    \includegraphics[width=\columnwidth]{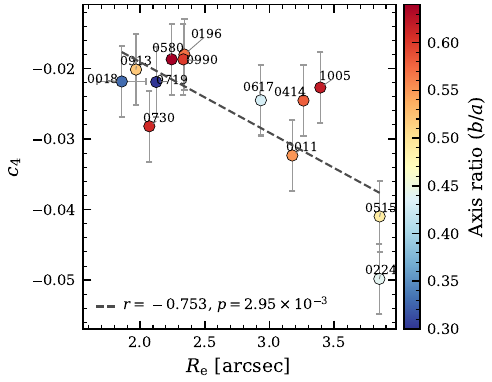}
    \caption{The observed anticorrelation between $c_4$ and effective radius ($R_{\rm e}$). Larger galaxies exhibit stronger boxy distortions (more negative $c_4$). Error bars on $c_4$ show the typical measurement uncertainty $\sigma_{\rm typ} = 0.005$ from the Appendix~\ref{app:validation} validation; $R_{\rm e}$ error bars are from the ERO catalog.}
    \label{fig:c4_vs_reff}
\end{figure}

We performed a statistical analysis of the correlations between the boxiness parameter and other structural properties (see Appendix~\ref{app:correlations}). We observe a significant anticorrelation between $R_{\rm e}$ and $c_4$ (Pearson $r \approx -0.75$, $p = 0.003$; Spearman $r_s = -0.59$, $p = 0.035$), as shown in Fig.~\ref{fig:c4_vs_reff}. Since more negative $c_4$ corresponds to stronger boxiness, this anticorrelation means that boxiness and $R_{\rm e}$ are positively correlated ($R_{\rm e} \uparrow \rightarrow$ Boxiness $\uparrow$), the physically largest dwarfs exhibit the most severe boxy distortions.

We caution, however, that the Pearson coefficient is sensitive to the two most boxy galaxies in the sample, EDwC-0224 and EDwC-0515. Removing both reduces the Pearson coefficient to $r = -0.45$ ($p = 0.17$), while the Spearman rank coefficient (less sensitive to outliers) gives $r_s = -0.59$ ($p = 0.035$) for the full sample. The trend is therefore real but should be interpreted with caution given the small sample size. We note that these same two galaxies also drive the tentative colour trend discussed below.

Boxiness acts as an independent structural feature rather than a proxy for profile shape: there is no statistically significant correlation between $c_4$ and concentration ($p = 0.59$) or the Sérsic index $n$ ($p = 0.75$).
This lack of correlation implies that the boxy morphology is not simply a result of how ``peaky'' or concentrated the galaxy light profile is, but rather represents a distinct non-axisymmetric component.

\subsection{Colour--shape correlation}
\label{sect:color_shape}

\begin{figure}
    \centering
    \includegraphics[width=\columnwidth]{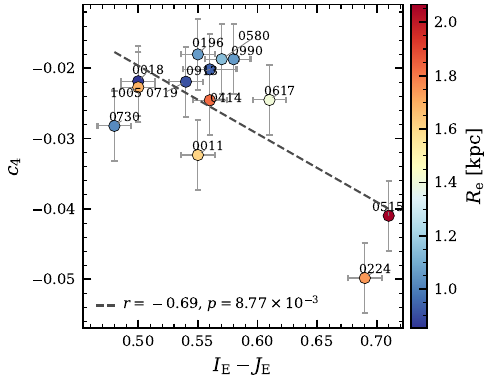}
    \caption{Tentative trend between \textit{Euclid} $\IE - \JE$ colour and boxiness $c_4$. Redder galaxies tend to exhibit stronger boxy distortions (more negative $c_4$), though this trend is driven by two galaxies at the extreme boxy end. Error bars on $c_4$ show the typical measurement uncertainty $\sigma_{\rm typ} = 0.005$ from the Appendix~\ref{app:validation} validation; colour error bars are propagated from the ERO catalog photometric uncertainties.}
    \label{fig:c4_vs_color}
\end{figure}

We observe a tentative trend between $c_4$ and galaxy colour ($\IE - \JE$) in Fig.~\ref{fig:c4_vs_color}: galaxies with more negative $c_4$ (more boxy) tend to be systematically redder. We note, however, that this trend is largely driven by two galaxies at the extreme boxy end of the distribution and should be interpreted with care. Although the colour variation is small and all targets lie on the cluster red sequence, this tentative correlation suggests that the boxiest dwarfs may host the most evolved stellar populations, consistent with these being the oldest, most tidally processed systems. The physical interpretation is discussed further in Sect.~\ref{sect:interpretation}.

\begin{figure*}
    \centering
    \includegraphics[width=0.8\textwidth]{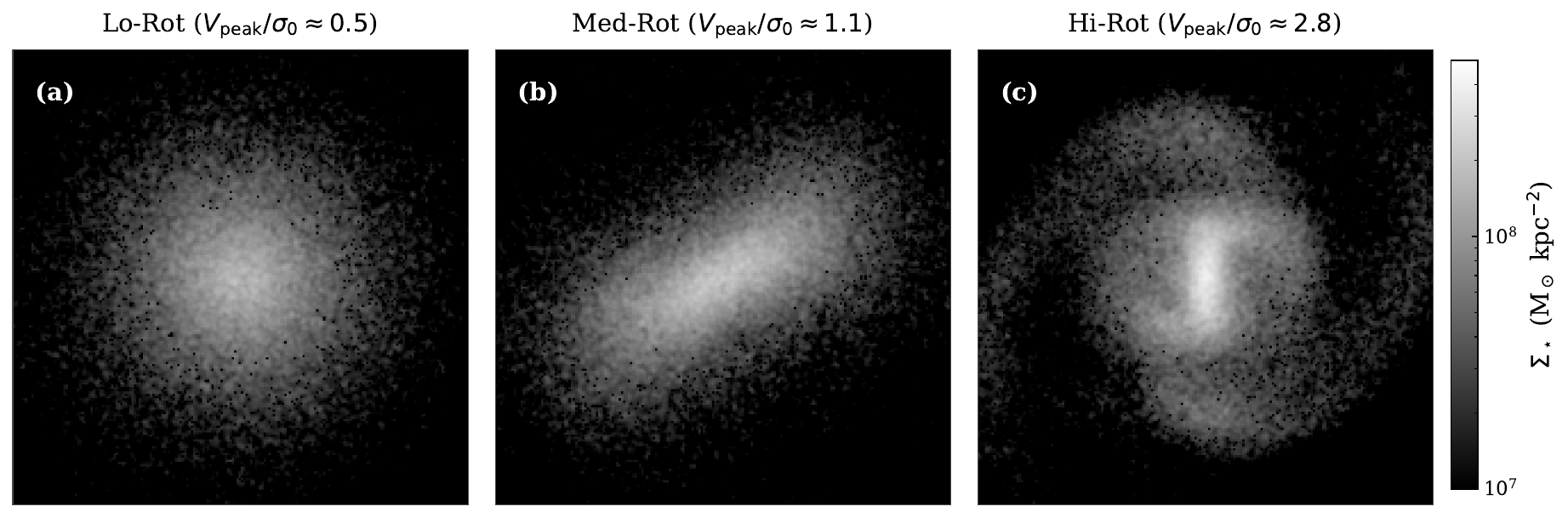}
    \caption{Projected stellar surface density at $t = 3.7$~Gyr (second pericentric passage) for the three rotation models of S21 on the strongly plunging Orbit~S1. Only the Med-Rot model develops the diffuse peanut structure consistent with our observed boxy dwarfs.}
    \label{fig:smith2021_surfdens}
\end{figure*}

\subsection{Observational constraints}
Our analysis of the 13 boxy dwarf galaxies in the Perseus cluster yields three specific observational constraints. Any physical model for their formation must simultaneously account for:

\begin{enumerate}
    \item Morphology: The galaxies exhibit significant boxy isophotes ($c_4 < -0.0175$) but lack any visual evidence of stellar disks or spiral arms, even at the diffraction-limited resolution of \textit{Euclid} \IE. They do not show the morphology of classical bars (which are features of spiral systems) or the boxy-peanut (B/P) bulge morphologies seen in massive disk galaxies; rather, their boxy distortion pervades the entire stellar body.

    \item Colour: Their integrated colours ($\IE - \JE \approx 0.57$) place them on the red sequence, implying they are gas-poor and fully quenched, as shown in the right panel of Fig.~\ref{fig:vis_nir_context}.

    \item Size--shape correlation: We observe a significant anticorrelation between $c_4$ and $R_{\rm e}$ (Pearson $r \approx -0.75$, $p=0.003$; $R_{\rm e} \uparrow \rightarrow$ Boxiness $\uparrow$).
\end{enumerate}

\section{Physical interpretation: the origin of boxiness}
\label{sect:interpretation}
\subsection{Tidal transformation and peanut formation}
We propose that the primary formation mechanism for these galaxies is the tidal transformation of progenitors with moderate rotational support. Throughout this section we develop this interpretation using a single simulated dwarf, the `Med-Rot' model ($V_{\rm peak}/\sigma_0 \approx 1.1$) on the strongly plunging `Orbit~S1' (pericentre $\approx 250$~kpc) from S21; the implications of relying on a single simulated archetype are discussed below. These are purely gravitational $N$-body simulations in which a model dwarf consisting of an NFW dark matter halo and a thick exponential stellar disk is subjected to the time-varying tidal field of a cluster potential; no gas physics or star formation is included, so the transformation is driven entirely by gravitational tidal forces. We use three snapshots ($t = 2.3$, 3.7, and 6.8~Gyr, corresponding to one, two, and four pericentric passages respectively); the morphological evolution between them is analysed in Sect.~\ref{sect:timescales}. The outcome of tidal transformation is sensitive to the initial dynamical state: while cold `Hi-Rot' disks develop bar-like structures and spiral arms, and hot `Lo-Rot' spheroids remain featureless, Med-Rot progenitors form diffuse, peanut-shaped inner structures under strong tidal forcing (Fig.~\ref{fig:smith2021_surfdens}). This transformation is characterised by differential stripping: while the dark matter halo is efficiently stripped, the stellar component remains largely intact, dynamically heated into an elongated peanut structure that survives as a recognisable feature.

\begin{figure*}
    \centering
    \includegraphics[width=0.6\textwidth]{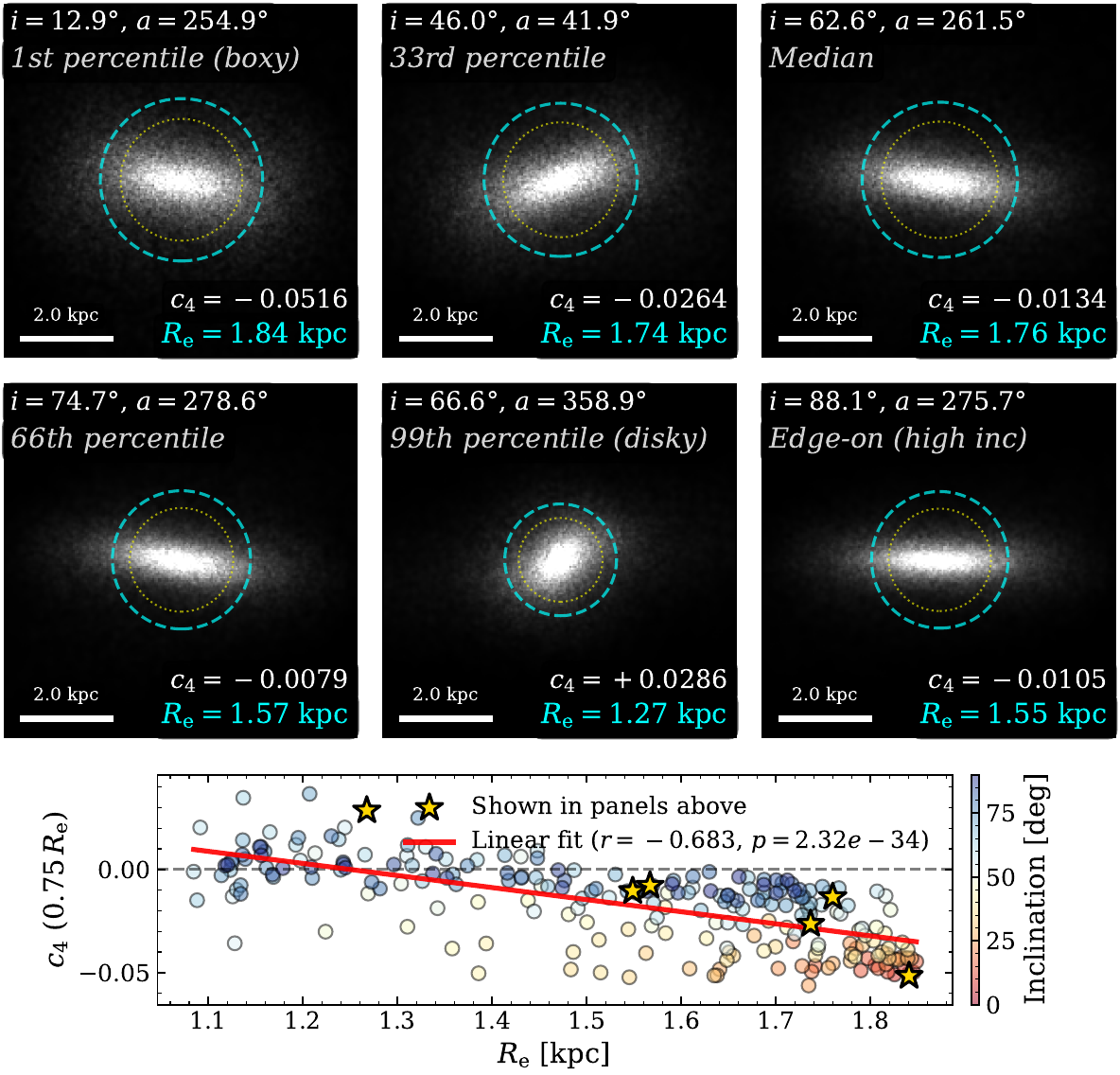}
    \caption{Simulated \textit{Euclid} \IE observations of the tidally transformed Med-Rot dwarf model at $t=3.7$~Gyr (S21), shown at selected percentiles of the $c_4$ distribution across all viewing angles, plus an edge-on view. The top row spans the 1st percentile (most boxy, $c_4=-0.052$, near face-on) to the median; the bottom row continues to the 66th and 99th percentiles (most disky, $c_4=+0.029$) and an edge-on case. The bottom panel shows $c_4$ versus $R_{\rm e}$ for all viewing angles, coloured by inclination, with the six selected orientations marked. Near face-on views produce the boxiest, largest isophotes, while higher inclinations yield rounder, more compact morphologies.}
    \label{fig:Re_vs_c4_medrot}
\end{figure*}

This choice of progenitor is expected to be largely insensitive to galaxy mass for the resulting morphology, although the timescales of the transformation may vary. The morphological outcome in the S21 simulations is governed by the dimensionless ratio $V_{\rm peak}/\sigma_0$ (the balance between ordered rotation and random motion). This ratio is scale-free, so the same dynamical regime (Med-Rot with $V_{\rm peak}/\sigma_0 \approx 1.1$) can in principle be realised at any stellar mass, suggesting that the tidal peanut-forming mechanism could apply across our sample's luminosity range, provided the progenitor has the requisite rotational support and plunges deep into the cluster potential.

Astrophysically, such progenitors are plausible. Since the Med-Rot condition describes a pre-infall state, the most relevant constraints come from field dwarf kinematics. \citet{2017MNRAS.465.2420W} found that $\sim 80$\% of Local Volume dwarfs are classified as dispersion-supported, with the majority of isolated systems having $V/\sigma \lesssim 1.0$, while \citet{2014MNRAS.441..452K} measured $V/\sigma \approx 1.5$ for star-forming blue compact dwarfs. Together, these surveys show that field dwarfs span $V/\sigma \approx 0.5$--$1.5$, bracketing the Med-Rot value of $V_{\rm peak}/\sigma_0 \approx 1.1$. Integral field spectroscopy of cluster dEs provides complementary, though less direct, evidence: \citet{2013MNRAS.428.2980R} found a wide variety of kinematic properties among 12 Virgo dEs observed with SAURON, and \citet{2015ApJ...799..172T} measured $V_{\rm rot}/\sigma_{\rm e}$ spanning 0 to $\approx 1.5$ across 39 Virgo dEs, with the majority of fast rotators falling in the range 0.3--1.2. Although these are measurements of already-transformed cluster members rather than progenitors, the prevalence of moderate rotational support among present-day dEs demonstrates that the Med-Rot regime is well populated. Thus, the Med-Rot condition represents a plausible starting point for a dwarf galaxy entering the cluster environment.

We also note that the Med-Rot simulation has a stellar mass of $3 \times 10^9 M_{\odot}$, which is roughly an order of magnitude more massive than the typical dwarf in our sample ($M_{\IE} \approx -15$; $M_* \sim 10^8 M_{\odot}$). However, our boxy candidates occupy a narrow range in both luminosity and mean surface brightness (see Fig.~\ref{fig:vis_nir_context}), suggesting they form a photometrically homogeneous population that can be reasonably represented by a single structural template. This homogeneity is consistent with our interpretation: mock observations of the simulation show that viewing angle induces $\sim 1$~mag~arcsec$^{-2}$ scatter in $\langle \mu_{\IE,\rm e} \rangle$ at fixed total luminosity, comparable to the observed spread in our sample (Appendix~\ref{app:sb_scatter}). While this does not imply that our candidates are literally a single object viewed from different angles, it does support the use of a single simulation as a representative structural archetype. We note, however, that our single-archetype model cannot account for possible variations in intrinsic size or Sérsic index across the sample. If the boxy dwarfs span a range of intrinsic structural properties, the observed $c_4$--$R_{\rm e}$ relation would reflect both projection effects and intrinsic diversity; disentangling these contributions would require simulations spanning a broader range of progenitor parameters. Furthermore, the transformation depends on the internal phase-space structure (set by $V_{\rm peak}/\sigma_0$) rather than absolute mass. Of the three S21 rotation classes, only the Med-Rot model develops the diffuse peanut structure that matches our observations (Fig.~\ref{fig:smith2021_surfdens}); it therefore serves as an appropriate analogue for any dwarf progenitor with moderate rotational support, irrespective of luminosity.

The observed morphological and photometric properties of our sample are consistent with the tidal transformation of moderately rotation-supported progenitors. This scenario provides a natural explanation for our first two constraints:

\begin{enumerate}
    \item Boxiness without disks: The transition from a disk-like progenitor to a peanutty spheroid is a characteristic outcome of tidal transformation within a cluster environment. In systems characterised by moderate initial rotational support, tidal torques and dynamical heating induce a boxy isophotal signature. This morphological endpoint is distinct from that of highly rotation-supported systems, which develop bar-like structures and are more likely to retain detectable spiral arms even after significant environmental processing (S21). The absence of visible disks or spiral features in our sample is therefore consistent with progenitors having intermediate initial rotational support, specifically the Med-Rot regime ($V_{\rm peak}/\sigma_0 \approx 1.1$): high enough to develop peanut structures under tidal forcing, but not so high as to preserve coherent disk or spiral features.

    \item Quenching: The uniform red colours of our sample align with the environmental processing inherent to the tidal transformation model. While gravitational interactions drive the structural transformation toward a peanutty spheroid, the simultaneous action of ram pressure stripping efficiently removes the extended gas reservoir. This stripping halts star formation and ensures that the resulting remnants are quenched systems residing on the cluster red sequence.
\end{enumerate}

\subsection{The geometric origin of the size--shape relation}
We propose that the observed positive correlation between effective radius $R_{\rm e}$ and boxiness ($R_{\rm e} \uparrow \rightarrow \text{Boxiness} \uparrow$), equivalently the negative correlation between $R_{\rm e}$ and $c_4$, is best understood as a consequence of geometric projection. This interpretation supports the tidal transformation scenario, suggesting that the observed trend may largely result from viewing a population of tidally transformed, peanutty remnants at varying orientations.

\begin{figure*}
    \centering
    \includegraphics[width=0.8\textwidth]{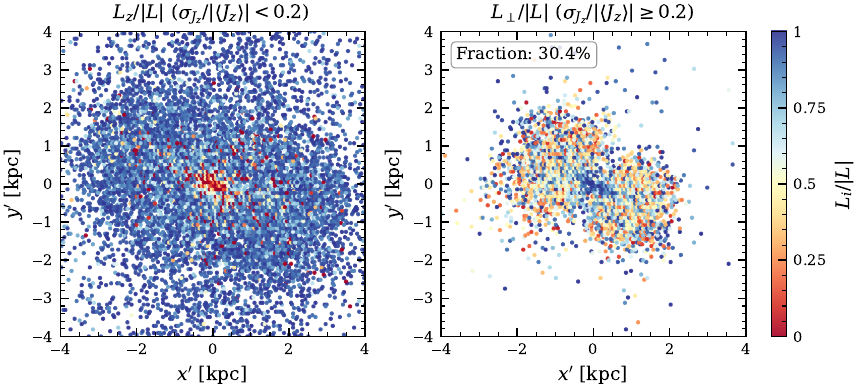}
    \caption{Separation of stable tube orbits (\emph{left}) and box orbits (\emph{right}, 30.4\% of stellar mass).
        Box orbits show enhanced perpendicular angular momentum components $|L_{\perp}| / |L|$.
        Note that the colour scale for the vertical component $L_z / |L|$ is clipped at 0.}
    \label{fig:orbital_structure}
\end{figure*}
We develop this argument in three steps: mock \textit{Euclid} observations show that projection alone reproduces the size--shape correlation (Sect.~\ref{subsect:mock_euclid}); a kinematic decomposition reveals box orbits in the inner regions, whose visibility depends on inclination (Sect.~\ref{sect:kinematics}); and snapshot evolution shows that extreme boxiness is a late-time feature (Sect.~\ref{sect:timescales}). Together these analyses connect the internal orbital structure of tidally transformed dwarfs to their observed isophotal shapes.

\subsubsection{Evidence from mock \textit{Euclid} observations}
\label{subsect:mock_euclid}
To test the projection hypothesis, we utilised the particle data from the Med-Rot dwarf simulation presented in S21, specifically subjected to the strongly plunging tidal orbit S1. We performed a multi-orientation analysis on the three evolutionary snapshots (2.3, 3.7, and 6.8~Gyr) to ensure our results are robust against time-dependent variations in the peanut structure.

To generate realistic mock observations comparable to our \textit{Euclid} data, we post-processed the simulation snapshots using the radiative transfer code \textsc{SKIRT} \citep{2020A&C....3100381C}. Since the original simulation is purely gravitational, we assigned each particle an old stellar population (8--12~Gyr) with sub-solar metallicities ($Z = 0.0004$--$0.007$), and generated SEDs using the BPASS models \citep{2017PASA...34...58E} with a \citet{2003PASP..115..763C} IMF and no dust. We produced mock \textit{Euclid} \IE-band images for 240 vantage points uniformly distributed on a sphere for each of the three snapshots, ensuring an unbiased sampling of possible orientations.

We then measured $R_{\rm e}$ and $c_4$ for each projection using the same methodology as our observational sample. Applying our observational selection criterion ($c_4 < -0.0175$) to the representative snapshot at $t=3.7$ Gyr, we find that 102 out of the 240 random orientations (42.5\%) result in a boxy classification. This substantial probability confirms that a single physical object, a tidally transformed peanut-dominated dwarf, will appear as a ``boxy dwarf'' from nearly half of all random viewing angles, while appearing as a standard, non-boxy spheroid from the remaining orientations. Figure~\ref{fig:Re_vs_c4_medrot} shows the resulting $R_{\rm e}$--$c_4$ distribution for the simulated projections, alongside representative mock observations at specific orientations. The images visually demonstrate that the ``broadside'' view yields the largest effective radius and strongest boxiness, while the ``end-on'' view results in a compact, rounder morphology.

Most interestingly, the simulation results reveal a clear anticorrelation between $c_4$ and $R_{\rm e}$ ($r \approx -0.68$) across the projections for all three snapshots. This trend is driven purely by the geometric orientation of the prolate peanut structure:
\begin{enumerate}
    \item {Face-on (Peanut Broadside):} When the galaxy is viewed face-on (perpendicular to the disk plane, $i < 60^\circ$), the elongated peanut structure is viewed broadside. In this orientation, the projected surface area is maximised (resulting in a large measured $R_{\rm e}$) and the full rectangular profile of the elongated triaxial structure is visible (resulting in strong boxiness, $c_4 \approx -0.05$).
    \item {Edge-on (Peanut End-on):} When the galaxy is viewed edge-on ($i > 60^\circ$) {and} the peanut structure is projected along the line of sight (end-on), the projected surface area is minimised (resulting in a small $R_{\rm e}$). In this configuration, the isophotes appear rounder ($c_4 \rightarrow 0$), diluting the boxiness signal.
\end{enumerate}
This result is consistent with the interpretation that our observed size--shape correlation is a geometric signature of a population of peanutty structures viewed at random orientations.
While low-inclination views ($i < 60^\circ$) consistently produce strongly negative $c_4$ values, high-inclination orientations span a range of boxiness driven by azimuthal angle: orientations where the major axis of the peanut structure lies perpendicular to the line of sight preserve a boxy signature, whereas foreshortened views of the elongated structure yield rounder isophotes.

\begin{figure}
    \centering
    \includegraphics[width=\columnwidth]{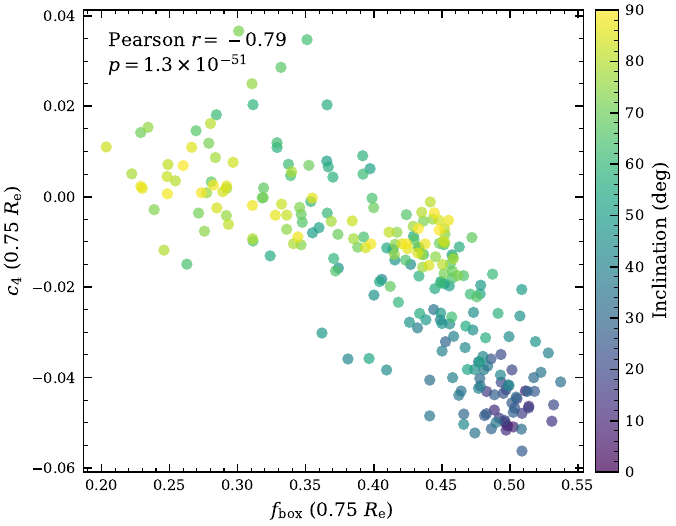}
    \caption{Correlation between the morphological isophote coefficient $c_4$ (more negative values indicate boxier isophotes) and the dynamical box-orbit light fraction $f_{\rm box}$ (the projected fraction of stellar light contributed by particles on box orbits at $0.75\,R_{\rm e}$) across 240 viewing angles of the Med-Rot model on Orbit~S1, coloured by inclination.}
    \label{fig:c4_fbox}
\end{figure}

\subsubsection{Intrinsic orbital structure and the geometric modulation of boxy signatures}
\label{sect:kinematics}

The preceding analysis demonstrates that viewing geometry can reproduce the size--shape correlation. We now turn to the underlying question: what is the physical origin of the boxy isophotes themselves? Specifically, we investigate whether a distinct orbital population is responsible for the boxy signature and how its visibility depends on viewing geometry. To address this, we performed a kinematic analysis of the Med-Rot model at $t = 3.7$ Gyr.

Stellar orbits within a triaxial potential are broadly classified into box and tube families, which are primarily distinguished by their angular momentum characteristics: while tube orbits maintain a stable, non-zero angular momentum, box orbits possess a mean angular momentum of zero, resulting in frequent sign changes and significant instantaneous fluctuations in the $z$-component $L_z$ \citep[see e.g.][]{Bovy2024}. To exploit this diagnostic, we used three closely-spaced simulation outputs separated by 25~Myr at each evolutionary epoch (distinct from the three widely-spaced evolutionary snapshots at 2.3, 3.7, and 6.8~Gyr discussed in Sect.~\ref{subsect:mock_euclid}), which provide the short-term temporal baseline needed to measure $L_z$ variability within individual particle trajectories.

We mapped the normalised variability of the $z$-component of angular momentum, defined as $\sigma_{L_z} / |\langle L_z \rangle|$, where $\sigma_{L_z}$ is the standard deviation of $L_z$ across the three closely-spaced outputs and $\langle L_z \rangle$ is its mean over the same interval, to distinguish between these populations. Particles identified as boxy orbits exhibit high variability and frequent sign reversals, confirming their association with the triaxial peanut structure (see Appendix~\ref{app:spatial_diagnostics}). Following this diagnostic, we adopted a variability threshold $\sigma_{L_z}/|\langle L_z \rangle| = 0.2$ (i.e., particles whose $L_z$ is stable to within 20\% of its mean over the sampled interval are classified as tubes, the rest as box orbits) to separate the particle population into two distinct categories; this value sits at the natural minimum between the coherently rotating and high-variability populations visible in the spatial maps of Appendix~\ref{app:spatial_diagnostics}. Particles with variability below this threshold are classified as stable tube orbits, characterised by a high and consistent $L_z / |L|$ ratio that indicates coherent circulation. Conversely, particles exceeding this threshold exhibit high $L_{\perp} / |L|$ values, where $L_\perp = \sqrt{L_x^2 + L_y^2}$ is the angular momentum component perpendicular to the galaxy's net angular momentum axis $\hat{z}$, confirming they are non-radial boxy orbits supported by the triaxial potential. This is shown in Fig.~\ref{fig:orbital_structure}. We note that the angular momentum of the peanut structure has a significant component perpendicular to the net angular momentum of the galaxy, which is dominated by the coherently circulating tube orbits.

We quantified how this intrinsic orbital composition manifests in projection by measuring the box orbit light fraction, $f_{\rm box}$, for each of the 240 viewing angles used in Sect.~\ref{subsect:mock_euclid}. For each orientation, we computed the projected surface density of box orbits, $\Sigma_{\rm box}$, relative to the total surface density, $\Sigma_{\rm total}$, extracting the value at $0.75\,R_{\rm e}$. We note that face-on is defined as the orientation parallel to the net angular momentum vector of the stellar component.

To directly test the link between orbital structure and isophotal shape, we compared the $c_4$ coefficient measured from the synthetic images to the corresponding $f_{\rm box}$ value at each viewing angle. The results, presented in Fig.~\ref{fig:c4_fbox}, reveal a strong anti-correlation between $c_4$ and $f_{\rm box}$ (Pearson $r = -0.79$, $p < 10^{-50}$), distinct from the observed $c_4$--$R_{\rm e}$ correlation discussed in Sect.~\ref{sect:correlations}. Orientations with higher projected box orbit fractions produce more negative $c_4$ values (boxier isophotes), while views dominated by tube orbits yield rounder profiles.

This correlation quantifies the geometric dilution of the boxy signature with viewing angle. In face-on orientations, the central boxy component is visible without dilution from material along the line of sight, yielding $f_{\rm box} \sim 0.5$ and maximally boxy isophotes ($c_4 \approx -0.05$). At high inclinations, tube orbits from the vertically heated thick disk project into the same region as the peanut structure, diluting both the box orbit fraction ($f_{\rm box} \sim 0.2$--0.3) and the boxy signature ($c_4 \approx 0$). The scatter in the $c_4$--$f_{\rm box}$ relation arises partly from azimuthal dependence: at fixed inclination, the observed $c_4$ varies with the orientation of the peanut structure relative to the line of sight. Accounting for this azimuthal effect tightens the correlation from $r = -0.79$ to $r = -0.91$. An example illustrating the dependence of $f_{\rm box}$ on inclination and radius is presented in Appendix~\ref{app:fbox_profiles}, together with a discussion of the $c_4$--$f_{\rm box}$ scatter in Appendix~\ref{app:azimuth_dependence}.

These findings directly connect the orbital structure to the observed size--shape correlation. Because box orbits are spatially concentrated in the elongated peanut region while tube orbits dominate the more spheroidal outer envelope, viewing geometry simultaneously modulates both the apparent size and boxiness. In face-on orientations, the full spatial extent of the elongated peanut structure is visible, maximising the measured effective radius; simultaneously, the box orbit population dominates the projected light at $0.75\,R_{\rm e}$, producing strong boxy isophotes. At high inclinations, the outcome depends on azimuthal angle: when the major axis of the peanut structure lies along the line of sight, both the apparent size and boxiness are reduced as tube orbits dilute the signal. However, when the major axis of the elongated structure is perpendicular to the line of sight even at high inclination, the rectangular profile of the inner peanut remains visible, preserving the boxy signature. This coupled geometric effect, where face-on viewing maximises both projected size and box orbit visibility, naturally produces the observed negative correlation between $R_{\rm e}$ and $c_4$. These findings suggest that the 13 boxy dwarfs identified in our Perseus sample likely represent the orientation-selected tail of a larger population of tidally transformed, peanutty systems. This lends support to the tidal transformation scenario, as the orientation-dependent coupling between apparent size and boxiness is a natural prediction of peanut-dominated remnants, thereby addressing our third observational constraint (Sect.~\ref{sect:results}).

\subsubsection{Timescales of morphological transformation}
\label{sect:timescales}
\begin{figure}
    \centering
    \includegraphics[width=\columnwidth]{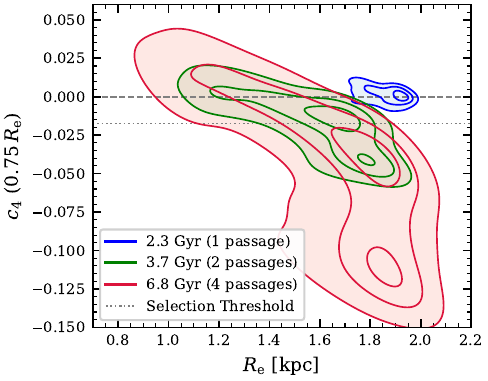}
    \caption{Evolutionary track of a tidally transformed dwarf (Med-Rot, Orbit S1) in the size--shape plane. Contours show 240 random viewing angles at three epochs: 2.3~Gyr (blue, first passage), 3.7~Gyr (green, peanut emergence), and 6.8~Gyr (red, extreme boxiness after four passages).}
    \label{fig:evolution_contours}
\end{figure}

The preceding analyses were performed at a single evolutionary snapshot ($t = 3.7$ Gyr). However, the inner peanut structure itself evolves under continued tidal forcing. We now examine how the morphology changes over multiple pericentric passages. Figure~\ref{fig:evolution_contours} illustrates the trajectory of the Med-Rot progenitor in a strongly plunging orbit in the size--shape plane across three distinct epochs. We find that tidal transformation drives a specific morphological sequence (values reported below are averaged over all viewing angles):

\begin{enumerate}
    \item Early interaction ($t=2.3$ Gyr): After the first passage, the galaxy retains a relatively large effective radius ($\langle R_{\rm e} \rangle \approx 1.84$ kpc) with little variation among viewing angles, and exhibits negligible boxiness ($\langle c_4 \rangle \approx 0$), appearing round or slightly disky.
    \item Intermediate processing ($t=3.7$ Gyr): Following the second passage, the effective radius decreases ($\langle R_{\rm e} \rangle \approx 1.56$ kpc) as tidal transformation leads to peanut formation. This produces moderate boxiness ($\langle c_4 \rangle \approx -0.017$), and the variation among viewing perspectives begins to intensify.
    \item Late-Stage ($t=6.8$ Gyr): After four passages\footnote{The last two pericentric passages are not as plunging.}, the boxiness intensifies significantly ($\langle c_4 \rangle \approx -0.051$) as the inner peanut becomes more pronounced and dominates the potential. The mean effective radius is $\langle R_{\rm e} \rangle \approx 1.60$ kpc, but the spread among vantage points is now substantial, reflecting the strong orientation dependence of the peanut-dominated morphology.
\end{enumerate}

This evolutionary track implies that the boxy morphology is a late-time feature that emerges only after repeated strong tidal shocks. Consequently, the most extreme boxy dwarfs in our sample are likely the oldest, most processed survivors in the cluster. The tentative colour--shape correlation noted in Sect.~\ref{sect:color_shape} is consistent with this evolutionary picture: if more extreme boxiness ($c_4 \ll -0.0175$) signals older, more tidally processed systems, these would have had longer to exhaust their star formation and thus appear redder. We caution, however, that this trend is driven by only two galaxies at the extreme end of the $c_4$ distribution and requires confirmation with a larger sample.

\subsection{Evidence against dry mergers}

\begin{table*}
    \centering
    \caption{Consistency of formation scenarios with observational and environmental constraints.}
    \label{tab:scenarios}
    \begin{tabular}{lcc}
        \hline
        {Constraint} & {Tidal Transformation} & {Dry Merger} \\
        \hline
        {Boxy Shape} & Yes (Peanut) & Yes (Violent Relaxation) \\
        {No Thin Disk} & Yes (Progenitor was thick/heated) & Yes (Scrambled momentum) \\
        {Red Colour} & Yes (Env. Quenching) & Yes (Dry Progenitors) \\
        {Size-Shape Trend} & {Yes (Projection)} & Unclear \\
        {Elongation ($b/a$)} & {Yes (Peanut geometry)} & {No (Too round)} \\
        {Environment} & {Favoured} (Cluster Potential) & {Tension} (Suppressed) \\
        \hline
    \end{tabular}
\end{table*}

While the tidal transformation of Med-Rot progenitors explains the data well, we must also consider dry mergers as an alternative channel. In massive galaxies, dry (gas-poor) mergers consistently produce boxy, pressure-supported remnants \citep{2006MNRAS.372..839N, 2014ApJ...787..102C}.

The geometric origin of boxiness differs between the two scenarios. In dry mergers, boxy isophotes typically arise from the projection of a triaxial or prolate spheroid viewed edge-on. In the tidal transformation scenario of S21, boxiness instead arises from a tidally induced peanut structure viewed face-on. Thus, the orientation dependence is inverted: the peanut-shaped remnant appears boxiest when the system is viewed face-on (revealing the full rectangular extent of the elongated structure), whereas an edge-on view results in a rounder projection where the boxy signature is lost.

We identify two primary lines of evidence that argue against the merger scenario for our sample. First, the axis ratios of our sample are inconsistent with merger remnants. While 1:1 mass ratio dry mergers can produce flattened remnants, simulations show these rarely achieve axis ratios lower than $b/a \sim 0.65$ \citep{2006MNRAS.372..839N}. Our sample exhibits a mean axis ratio of $b/a \approx 0.55$, which is significantly more elongated than expected for merger products but consistent with a peanut-shaped triaxial structure viewed broadside. We note that these simulations focus on equal-mass mergers with specific orbital configurations and do not comprehensively survey all merger parameter space.

Furthermore, the high velocity dispersion of the Perseus cluster core ($\sigma_{\rm cl} \sim 1300$ km s$^{-1}$; \citealt{1983AJ.....88..697K}) strongly suppresses the cross-section for slow, gravitationally bound mergers: typical dwarf galaxy escape velocities are of order a few tens of km~s$^{-1}$, so relative encounter velocities in the cluster ($v_{\rm rel} \sim \sqrt{2}\,\sigma_{\rm cl}$) vastly exceed the escape speed, and encounters are too fast for gravitational capture.
Therefore, if these are merger remnants, they likely formed in the group environment prior to infall, a process known as ``pre-processing'' \citep{2004cgpc.symp..277M, 2017MNRAS.468.4625J, 2018ApJ...866...78H, 2020MNRAS.497.1904B}.
Given the dynamical constraints and morphological inconsistencies, we favour the tidal transformation scenario. Table~\ref{tab:scenarios} summarises the consistency of both scenarios with our observational constraints.

Taken together, these results support a coherent formation picture for the boxy dwarf population in Perseus. Tidal transformation of moderately rotating progenitors produces peanut-dominated remnants whose inner regions are shaped by box orbits. The observed size--shape correlation arises naturally from the projection of these prolate structures: face-on views reveal the full extent and rectangular profile of the elongated peanut structure, while edge-on views yield smaller, rounder isophotes as tube orbits dilute the boxy signal. This geometric effect implies that the 13 boxy dwarfs identified in our sample likely represent an orientation-selected subset of a larger population of environmentally transformed, peanutty systems, having traversed the cluster core on plunging orbits. The absence of thin disks, the quenched stellar populations, and the characteristic axis ratios are all consistent with this scenario, while the dynamical environment of Perseus disfavors dry mergers as a competing channel.

Direct kinematic confirmation of this picture will require integral field spectroscopy. The simulations predict that the angular momentum is carried by the outer tube orbits, while the inner peanut region (where box orbits average to zero net rotation, Sect.~\ref{sect:kinematics}) is dispersion-supported; the corresponding signature in resolved kinematic data is a $V/\sigma$ profile that declines from the rotating outer envelope toward the dispersion-dominated centre. At the distance of Perseus such measurements are observationally challenging, but boxy dwarf candidates in nearer clusters or in the Local Volume offer more accessible targets for an analogous test.

\section{Summary and conclusions}
\label{sect:summary}

We have presented a morphological analysis of dwarf galaxies in the Perseus cluster using ERO data. By developing a novel methodology for measuring isophotal shapes in low surface brightness systems and comparing our observations with gravitational $N$-body simulations of tidally transformed dwarfs, we have identified a population of boxy early-type dwarfs whose properties are best explained by tidal transformation. Our principal results are as follows:

\begin{enumerate}
    \item {Novel methodology:} We developed a cumulative light fraction approach for measuring the $c_4$ Fourier coefficient that is specifically designed for low surface brightness systems. Rather than extracting contours at fixed surface brightness levels, we measure isophotes at 20 equally-spaced flux fractions (0.2--0.9) and compute $c_4 = (2/N)\sum_i \delta(\theta_i) \cos(4\theta_i)$ from normalised deviations relative to fitted ellipses. We report $c_4$ at $0.75 R_{\rm e}$. Validation against \textit{Euclid}-realistic synthetic galaxies matched to our sample properties (Appendix~\ref{app:validation}) yields noise-induced uncertainties of $\sigma(c_4) \approx 0.003$--$0.008$, confirming that our boxy detections are robust. This method is directly applicable to isophotal shape analysis in any low surface brightness regime.

    \item {Boxy dwarf population:} From the 1061 classified dEs in Perseus, we select a clean sample of 183 systems sufficiently resolved and free of contamination for reliable $c_4$ measurement. From this clean sample, we identify 13 dwarfs with significantly boxy isophotes ($c_4 < -0.0175$), confirmed through visual inspection, with values ranging from $-0.050$ to $-0.018$. These galaxies lack visible disk components or spiral structure at \textit{Euclid}'s diffraction-limited resolution and lie uniformly on the red sequence ($\IE - \JE \approx 0.57$), indicating quenched, gas-poor systems.

    \item {Size--shape anticorrelation:} We find a significant anticorrelation between $c_4$ and effective radius (Pearson $r = -0.75$, $p = 0.003$; Spearman $r_s = -0.59$, $p = 0.035$; $R_{\rm e} \uparrow \rightarrow$ Boxiness $\uparrow$): the physically largest dwarfs exhibit the most pronounced boxy distortions, with the trend driven in part by the two most boxy galaxies in the sample. The parameter $c_4$ shows no correlation with concentration or S\'ersic index ($p > 0.5$), demonstrating that boxiness represents a distinct non-axisymmetric structural component independent of the radial light profile shape. The boxy dwarfs also show no preferential spatial concentration within Perseus (K--S test $p = 0.65$ versus the wider dE sample). We also note a tentative colour--shape trend: boxier galaxies tend to be redder, though this trend is driven primarily by two galaxies at the most extreme end of the $c_4$ distribution and must be treated with caution.

    \item {Tidal transformation scenario:} Comparison with simulations from S21 strongly favours tidal transformation of moderately rotating progenitors ($V_{\rm peak}/\sigma_0 \approx 1.1$) over dry mergers. Mock \textit{Euclid} observations of tidally transformed Med-Rot models show that $\sim 42$\% of random viewing angles produce boxy classifications ($c_4 < -0.0175$), indicating that our observed sample represents an orientation-selected subset of a larger population. The simulation reproduces the observed size--shape correlation ($r \approx -0.68$) purely through projection effects: face-on orientations maximize both apparent size and boxiness by revealing the full rectangular profile of the tidally induced peanut structure, while edge-on views yield rounder, smaller appearances.

    \item {Orbital structure and the size--shape correlation:} Kinematic analysis of the simulations reveals that box orbits, characterised by high angular momentum variability and frequent $L_z$ sign changes, constitute $\sim 30$\% of the stellar mass and are spatially concentrated in the elongated peanut region, while tube orbits dominate the more spheroidal outer envelope. This spatial segregation directly explains the size--shape correlation: viewing geometry simultaneously modulates both apparent size and boxiness. Face-on orientations reveal the full extent of the elongated peanut structure (maximising $R_{\rm e}$) while box orbits dominate the projected light (maximising boxiness, $c_4 \approx -0.05$). At high inclinations with the peanut structure along the line of sight, both the apparent size and boxiness are reduced as tube orbits dilute the signal. We find a strong anti-correlation between $f_{\rm box}$ and $c_4$ across 240 viewing angles (Pearson $r = -0.79$, tightening to $r = -0.91$ when accounting for azimuthal dependence).

    \item {Evolutionary processing:} Analysis of simulation snapshots at three epochs (2.3, 3.7, and 6.8 Gyr) reveals that significant boxiness emerges after the second pericentric passage ($t = 3.7$ Gyr, $c_4 \approx -0.017$), intensifying with subsequent encounters to extreme values ($t = 6.8$ Gyr, $c_4 \approx -0.051$) after four passages. The most boxy dwarfs in our sample are therefore likely late-stage, heavily processed remnants that have survived multiple strong tidal shocks, having plunged through the cluster core on their orbits.

    \item {Dry mergers disfavoured:} We find dry mergers to be disfavored based on two primary lines of evidence: (1) the sample's mean axis ratio ($b/a \approx 0.55$) is more elongated than typical equal-mass merger remnants ($b/a \gtrsim 0.65$; \citealt{2006MNRAS.372..839N}), and (2) Perseus's high velocity dispersion ($\sigma_{\rm cl} \sim 1300$ km s$^{-1}$) suppresses the slow interactions required for mergers. Furthermore, the geometric origin of boxiness differs fundamentally between scenarios: in our tidal transformation model, boxiness arises from viewing a tidally induced peanut structure face-on, whereas in mergers it typically arises from viewing a triaxial spheroid edge-on.

\end{enumerate}

Our results demonstrate that environment-driven transformation in cluster cores can produce morphologies traditionally associated with mergers. Tidally transformed dwarfs with moderate rotation form diffuse, peanut-shaped inner structures that dominate the stellar light, providing an observational signature distinguishable through detailed isophotal analysis. The orientation-dependence of this signature means that only a fraction of tidally processed remnants appear boxy at any given time; correcting for inclination effects by dividing our 13 detections by the 42.5\% boxy fraction recovered from the simulations, and assuming the \citet{EROPerseusDGs} catalogue is representative of the cluster dwarf population, the parent population is likely $\sim 30$ galaxies in Perseus.

The strong size--shape correlation highlights the role of viewing geometry in interpreting morphological measurements of peanutty systems. More broadly, the connection between observed isophotal shapes and specific progenitor properties (here, moderate rotational support; \citealt{2017MNRAS.465.2420W, 2014MNRAS.441..452K}) demonstrates how detailed morphological analysis can constrain the dynamical histories of cluster dwarfs.

Future integral field spectroscopy will be essential for kinematic confirmation of the tidal transformation scenario, providing direct observational constraints on the orbital structure predicted by the simulations. On the theoretical side, dedicated simulations of dwarf galaxy tidal transformation in the mass range of our observed sample ($M_* \sim 10^8 M_{\odot}$) represent a natural next step; while our analysis suggests that rotational support rather than mass governs the morphological outcome, simulations matched to the observed luminosity range would enable direct quantitative comparison with the data. Equally valuable would be a systematic exploration of initial conditions spanning a wide range of rotational support ($V_{\rm peak}/\sigma_0$), progenitor sizes, and Sérsic indices, which would clarify how progenitor kinematics and structure map onto the full diversity of observed isophotal shapes and quantify how much of the observed $c_4$--$R_{\rm e}$ scatter is attributable to projection effects versus intrinsic structural variation. Extension of our methodology to other \textit{Euclid}-observed clusters will enable the first systematic census of tidally transformed dwarfs, constraining tidal transformation efficiency as a function of cluster mass, dynamical state, and clustercentric radius.

\section{Data availability}

The \textit{Euclid} data used in this work are available through the \textit{Euclid} Archive at \url{https://easotf.esac.esa.int/sas/}. The morphological measurements, analysis code, and galaxy cutouts are available on reasonable request to the authors.

\begin{acknowledgements}
AUK acknowledges support from the Belgian Federal Science Policy Office (BELSPO) via the PRODEX Programme of the European Space Agency (ESA) under contract number 4000143202. SDR acknowledges funding from the FWO. RS acknowledges financial support from FONDECYT Regular 2023 project No.\ 1230441 and also gratefully acknowledges financial support from ANID -- MILENIO NCN2024\_112. Co-funded by the European Union (MSCA EDUCADO, GA 101119830). Views and opinions expressed are however those of the author(s) only and do not necessarily reflect those of the European Union. Neither the European Union nor the granting authority can be held responsible for them.
\AckEC
\AckERO\cite{EROcite}
This research made use of \texttt{MTObjects}, \texttt{statmorph}, \texttt{Astropy}, \texttt{matplotlib}, and \texttt{NumPy}.
\end{acknowledgements}


\FloatBarrier
\bibliographystyle{aa}
\bibliography{bibliography,Euclid}

\FloatBarrier
\appendix

\section{\texttt{MTObjects} segment scoring: worked example}
\label{app:mtobjects_example}

\begin{table*}[htbp]
\centering
\caption{\texttt{MTObjects} analysis for EDwC-0011 in the \IE band. Of 82 detected segments, only one passes all selection criteria. A representative subset is shown illustrating the main rejection categories.}
\label{tab:mtobjects_vis}
\begin{tabular*}{\textwidth}{@{\extracolsep{\fill}}rcccccccl}
\hline\hline
\textbf{ID} & \textbf{$\IE$} & \textbf{$\langle\mu\rangle$} & \textbf{Area (arcsec$^2$)} & \textbf{Edge (\%)} & \textbf{Dist (px)} & \textbf{Score} & \textbf{Status} & \textbf{Rejection Reason} \\
\hline
7  & 25.26 & 25.49 &  0.33 &  0.00 & 106.52 & $-32.40$ & Rejected & Too faint, Too small \\
27 & 24.14 & 25.16 &  0.53 &  0.00 &  54.48 & $-16.18$ & Rejected & Too small \\
46 & 20.67 & 22.40 &  1.01 &  0.00 &  68.23 & $-20.46$ & Rejected & Too small \\
51 & 21.43 & 19.13 &  0.08 &  0.00 & 100.46 & $-30.51$ & Rejected & SB too high, Too small \\
61 & 19.76 & 24.48 & 17.98 &  0.00 &   1.92 &    0.40  & \textbf{MAIN} & --- \\
82 & 30.32 &  ---  &  0.00 & 57.74 & 162.52 & $-49.98$ & Rejected & Too faint, Too small \\
\hline\hline
\multicolumn{9}{p{\textwidth}}{\small \textbf{Note:} Object 61 is the only segment passing all selection criteria (surface brightness $\ge 22.0$~mag~arcsec$^{-2}$, magnitude 15.5--25.0, equivalent radius $>15$~pixels, axis ratio $<10$). With a single valid segment, the centrality and area scores are degenerate: Centrality~$= 0.000$, Area Score~$= 1.000$, Edge Factor~$= 1.000$, yielding Final Score~$= 0.400$.}
\end{tabular*}
\end{table*}

Table~\ref{tab:mtobjects_vis} illustrates the segment scoring procedure (Sect.~\ref{sect:methodology}) for a representative galaxy, EDwC-0011. Of 82 detected segments, only one passes all selection criteria.

\section{Validation of the isophotal shape measurement}
\label{app:validation}

We validated our $c_4$ measurement using synthetic galaxies with \textit{Euclid}-realistic noise properties matched to the observed boxy sample. The synthetic galaxies consist of a S\'ersic profile with angular intensity modulation following the standard isophote fitting parametrisation \citep{1999BaltA...8..535M}:
\begin{equation}
I(R,\Theta) = I(R) \, \left[1 + \delta \cos(4\Theta)\right] \,,
\label{eq:intensity_modulation}
\end{equation}
\noindent where $I(R)$ is the S\'ersic profile evaluated at elliptical radius $R$, $\Theta$ is the azimuthal angle measured from the major axis, and $\delta$ is the shape modulation amplitude (negative values produce boxy isophotes).

Each synthetic image is convolved with the \textit{Euclid} \IE PSF (Gaussian with FWHM~$= 0\farcs16$, corresponding to 1.6~pixels) and includes a physically motivated noise model: Poisson noise from both the source and sky background ($\mu_{\rm sky} = 22.3$~mag~arcsec$^{-2}$; \citealt{EROPerseusOverview}, Table~1), plus read noise ($3.2~e^{-}/$frame $\times \sqrt{16}$ frames). We fix $R_{\rm e} = 25$~pixels ($2\farcs5$), representative of the boxy sample (\texttt{statmorph} $R_{\rm e}$ range 19--39~pixels), S\'ersic index $n = 1.0$ (boxy sample median 1.07), and axis ratio $b/a = 0.4$ (representing the more elongated boxy dwarfs, sample range 0.30--0.95, median 0.55). The shape parameter $\delta$ spans ten values randomly drawn from $[-0.12, 0.05]$, bracketing the observed $c_4$ range. Peak signal-to-noise ratios of 30, 45, 55, 70, and 85 match the measured range of our boxy sample (median~53, measured via $5 \times 5$ median-filtered peak flux to exclude nuclei and globular clusters). For each of the 50 configurations ($10~\delta \times 5$~S/N), we generate five independent noise realisations, totalling 250 measurements.

\begin{figure*}
    \sidecaption
    \includegraphics[width=12cm]{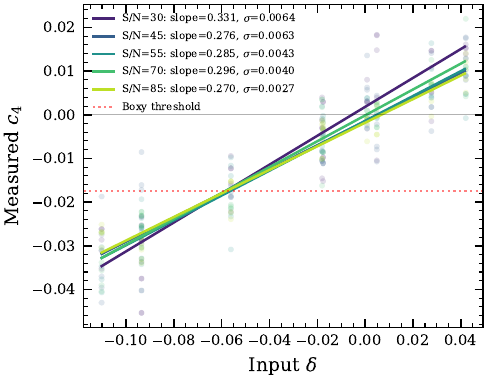}
    \caption{Validation of the $c_4$ measurement using \textit{Euclid}-realistic synthetic galaxies ($n = 1.0$, $R_{\rm e} = 25$~pix, $b/a = 0.4$). Measured $c_4$ versus input shape parameter $\delta$ at five S/N levels matched to the observed boxy sample. Solid lines show linear fits; the horizontal dashed line marks the boxy threshold $c_4 = -0.0175$.}
    \label{fig:validation_a4_c4}
\end{figure*}

Figure~\ref{fig:validation_a4_c4} shows the measured $c_4$ versus input $\delta$ at each S/N level. The relationship is linear with slope $\approx 0.28$ and scatter that decreases with increasing S/N. For galaxies in the boxy regime ($c_4 < -0.0175$), the noise-induced measurement scatter is $\sigma(c_4) \approx 0.003$--$0.008$ across the S/N range of our sample (Table~\ref{tab:validation_summary}). We note that this validation addresses noise-related uncertainties only, systematic effects from source contamination, segmentation artefacts, or centring errors are not captured by these synthetic tests.

\begin{table}
\centering
\caption{Validation results for $n = 1.0$, $R_{\rm e} = 25$~pix, $b/a = 0.4$.}
\label{tab:validation_summary}
\begin{tabular}{cccc}
\hline\hline
S/N & Slope & RMSE & $\sigma_{\rm boxy}$ \\
\hline
30 & 0.331 & 0.006 & 0.006 \\
45 & 0.276 & 0.006 & 0.008 \\
55 & 0.285 & 0.004 & 0.005 \\
70 & 0.296 & 0.004 & 0.004 \\
85 & 0.270 & 0.003 & 0.003 \\
\hline\hline
\multicolumn{4}{p{0.9\columnwidth}}{\small \textbf{Notes.} Slope: best-fit linear slope of $c_4$ vs.\ $\delta$. RMSE: root-mean-square scatter across all $\delta$ values. $\sigma_{\rm boxy}$: scatter for configurations with mean $c_4 < -0.0175$.}
\end{tabular}
\end{table}

\section{Distribution of \texorpdfstring{$c_4$}{c4} in the clean sample}
\label{app:c4_distribution}

\begin{figure}
    \centering
    \includegraphics[width=0.5\textwidth]{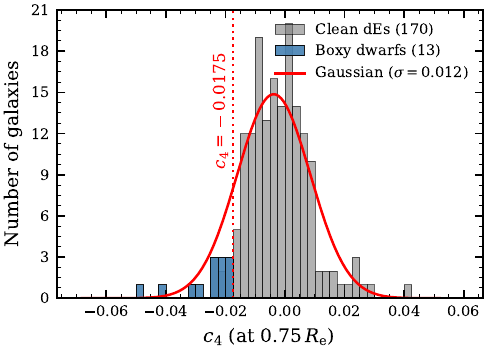}
    \caption{Distribution of $c_4$ for the 183 clean early-type dwarfs. Grey bars show the non-boxy population; blue bars highlight the 13 visually confirmed boxy dwarfs. The red curve is a Gaussian fit ($\sigma = 0.012$) to the full sample. The dotted line marks the boxy threshold $c_4 = -0.0175$.}
    \label{fig:c4_histogram}
\end{figure}

Figure~\ref{fig:c4_histogram} shows the $c_4$ distribution for the full clean sample of 183 early-type dwarfs. The distribution is approximately Gaussian with $\sigma = 0.012$, centred near zero. However, this observed width exceeds the noise-induced measurement uncertainty of $\sigma(c_4) \approx 0.003$--$0.008$ established by our matched validation (Appendix~\ref{app:validation}), indicating that a significant fraction of the observed scatter reflects genuine intrinsic variation in isophotal shape across the dE population. The 13 visually confirmed boxy dwarfs (blue) occupy the negative tail of this distribution, with $c_4$ values ($-0.050$ to $-0.018$) representing individually significant detections: each lies $3$--$15\sigma$ below zero given the per-galaxy measurement uncertainties.

\section{Correlations between boxiness and galaxy properties}
\label{app:correlations}

\begin{figure*}
    \sidecaption
    \includegraphics[width=12cm]{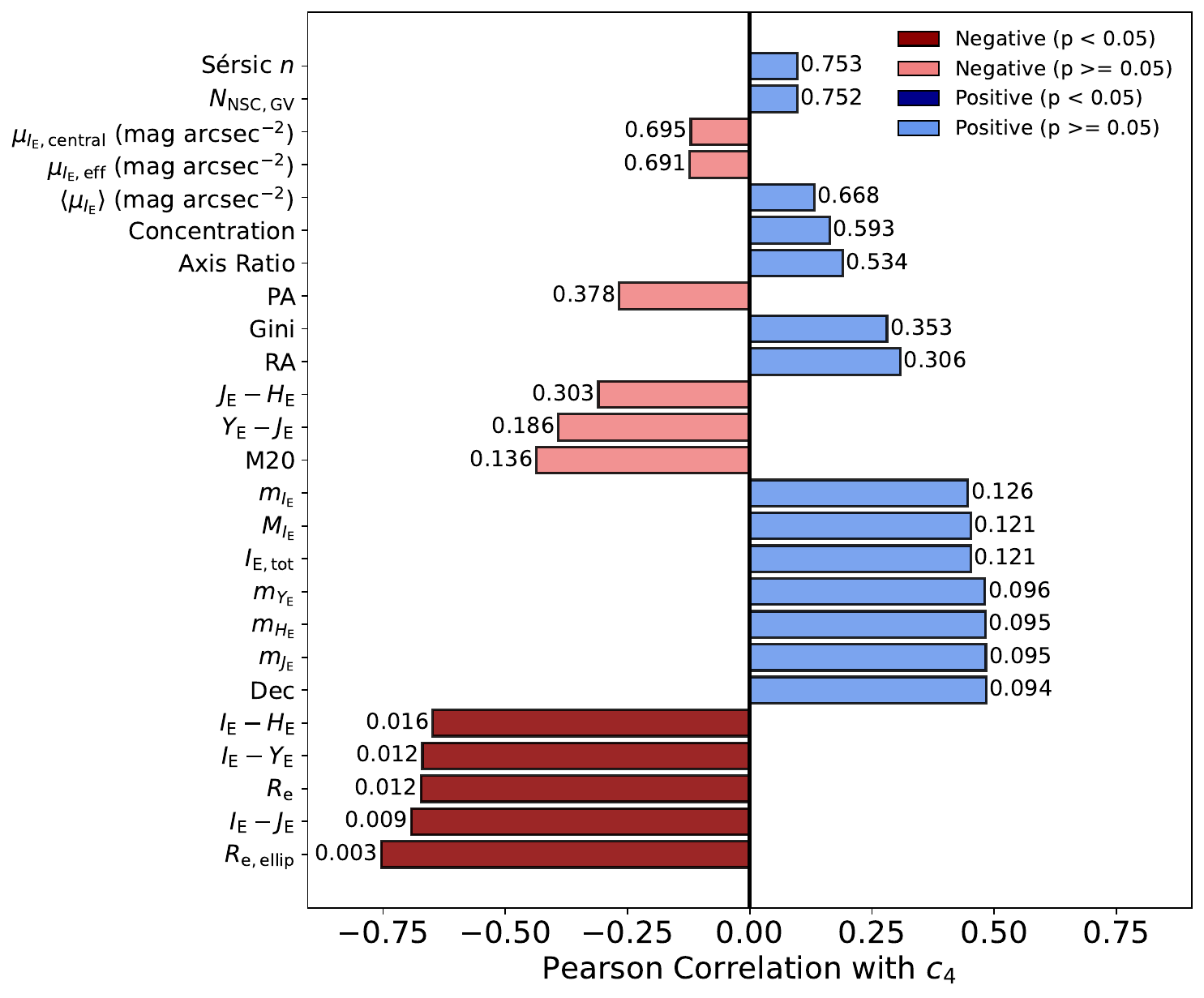}
    \caption{Correlations between the boxiness parameter $c_4$ and various galaxy properties for the 13 boxy Perseus dwarfs.}
    \label{fig:c4_correlations_csv}
\end{figure*}

Figure~\ref{fig:c4_correlations_csv} shows the full set of correlations between $c_4$ and structural properties of the Perseus dwarf sample. The main result (a significant anticorrelation between $c_4$ and $R_{\rm e}$, Pearson $r \approx -0.75$, $p = 0.003$) is discussed in the main text (Sect.~\ref{sect:correlations}). The remaining panels confirm the absence of statistically significant correlations with concentration ($p = 0.59$) and Sérsic index ($p = 0.75$), supporting the interpretation that boxiness is an independent structural feature.

\section{Orbital diagnostics: \texorpdfstring{$L_z$}{Lz} variability maps}
\label{app:spatial_diagnostics}

\begin{figure*}
    \sidecaption
    \includegraphics[width=12cm]{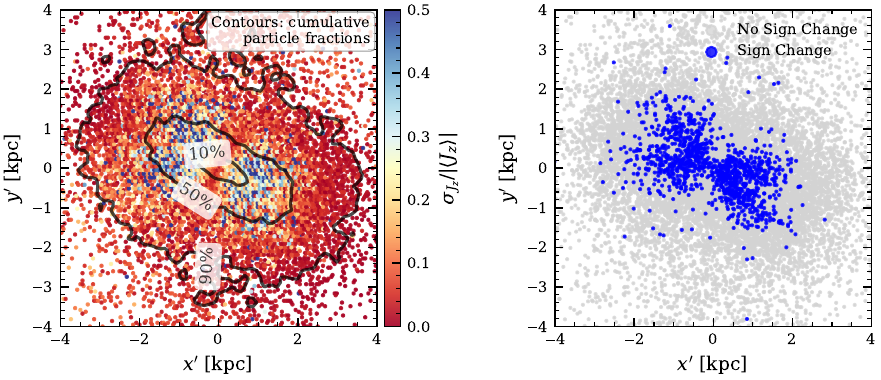}
    \caption{Orbital diagnostics at $t=3.7$ Gyr. \emph{Left}: normalised $L_z$ variability. \emph{Right}: particles with $L_z$ sign changes. High-variability regions coincide with box orbits in the inner peanut structure.}
    \label{fig:spatial_diagnostics}
\end{figure*}

Figure~\ref{fig:spatial_diagnostics} shows the spatial distribution of the normalised $L_z$ variability ($\sigma_{L_z}/|\langle L_z \rangle|$) and sign-change fraction across the Med-Rot model at $t=3.7$~Gyr. High-variability regions coincide with the inner peanut structure, confirming that box orbits are preferentially located in the elongated central component. This diagnostic motivates the variability threshold of 0.2 used in Sect.~\ref{sect:kinematics} to separate tube from box orbits.

\section{Box orbit fraction profiles}
\label{app:fbox_profiles}

\begin{figure*}
    \sidecaption
    \includegraphics[width=12cm]{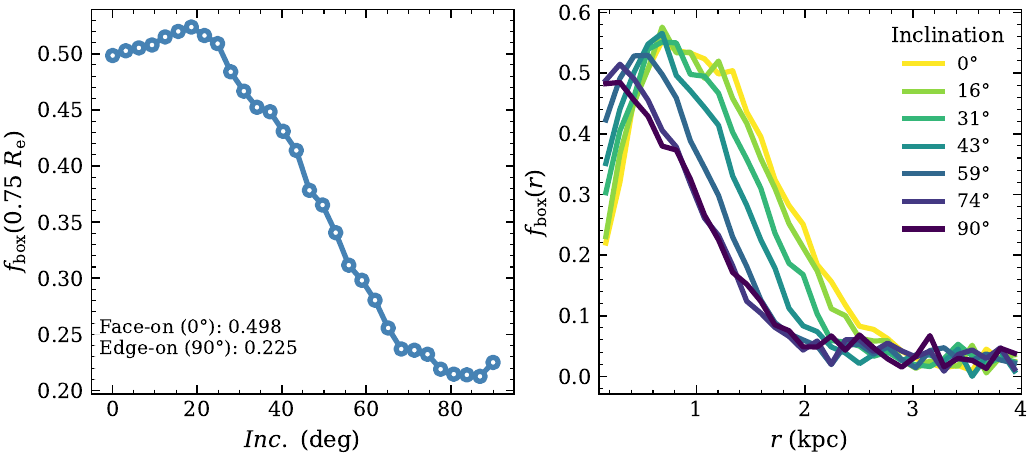}
    \caption{Projected box orbit fraction versus viewing angle. \emph{Left}: one-dimensional slice showing $f_{\rm box}$ evaluated at a single radius ($0.75\,R_{\rm e}$) as a function of inclination. \emph{Right}: the full radial profiles $f_{\rm box}(r)$ at several fixed inclinations, of which the left panel is the vertical slice at $0.75\,R_{\rm e}$. Colours encode inclination (light to dark as inclination increases from face-on to edge-on).}
    \label{fig:fbox_inclination}
\end{figure*}

Figure~\ref{fig:fbox_inclination} shows the dependence of the projected box orbit fraction $f_{\rm box}(r) = \Sigma_{\rm box}(r) / \Sigma_{\rm total}(r)$ on viewing angle and galactocentric radius. For this figure, we sampled 30 inclinations uniformly from face-on to edge-on at a fixed azimuth. The left panel demonstrates the systematic decline in $f_{\rm box}$ from face-on ($\sim$0.50) to edge-on ($\sim$0.23) at $0.75\,R_{\rm e}$, a $\sim$55\% reduction reflecting geometric dilution as tube orbits from the thick disk project into the same region as the peanut structure at high inclinations. The right panel shows radial profiles at selected inclinations, revealing that box orbits dominate the inner regions at all viewing angles but their relative contribution to the projected light is most prominent in face-on views.

\section{Scatter in the \texorpdfstring{$c_4$--$f_{\rm box}$}{c4--fbox} relation}
\label{app:azimuth_dependence}

\begin{figure*}
\sidecaption
\includegraphics[width=12cm]{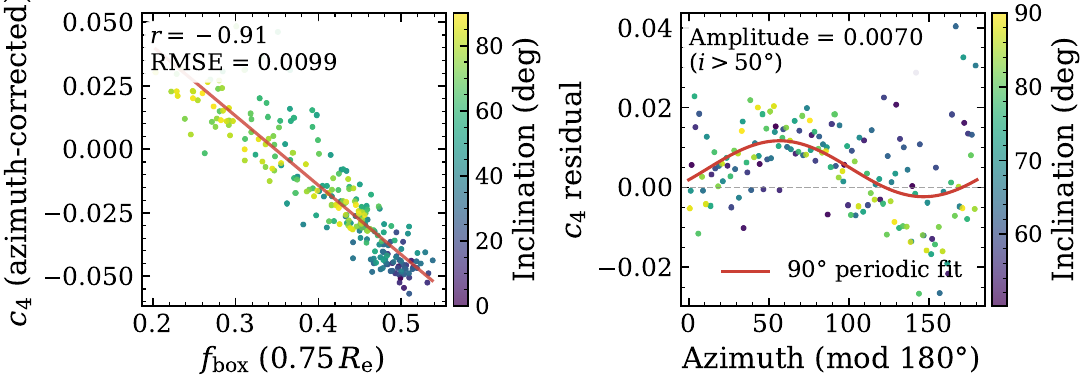}
\caption{Azimuthal dependence of the $c_4$--$f_{\rm box}$ relation. \emph{Left}: correlation after removing azimuthal terms (Eq.~\ref{eq:c4_resid}), coloured by inclination. \emph{Right}: $c_4$ residuals versus azimuth at high inclinations ($i > 50^{\circ}$)}
\label{fig:azimuth_dependence}
\end{figure*}

The scatter in the $c_4$--$f_{\rm box}$ relation arises partly from the azimuthal orientation of the peanut structure relative to the observer (Fig.~\ref{fig:azimuth_dependence}). To quantify this, we fit a model that includes both $f_{\rm box}$ and azimuthal terms,
\begin{equation}
c_4 = a \, f_{\rm box} + \big(b + d \tfrac{i}{90^{\circ}}\big) \sin(2\phi) + \big(c + e \tfrac{i}{90^{\circ}}\big) \cos(2\phi) + f \,,
\label{eq:c4_model}
\end{equation}
where $\phi$ is the azimuth angle and $i$ is the inclination. The interaction terms (proportional to $i$) account for the azimuthal effect strengthening at higher inclinations.
To remove the azimuthal dependence, we subtract only the azimuth-related terms from the observed $c_4$,
\begin{equation}
c_4^{\rm corr} = c_4 - \big(b + d \tfrac{i}{90^{\circ}}\big) \sin(2\phi) - \big(c + e \tfrac{i}{90^{\circ}}\big) \cos(2\phi) \,,
\label{eq:c4_resid}
\end{equation}
leaving $c_4^{\rm corr} \approx a \, f_{\rm box} + f + \epsilon$, where $\epsilon$ is the residual scatter.
After removing the $f_{\rm box}$ trend, we fit the residuals to isolate the azimuthal dependence,
\begin{equation}
c_4^{\rm resid} = A \sin(2\phi) + B \cos(2\phi) + C \,,
\end{equation}
with amplitude $\sqrt{A^2 + B^2} \approx 0.008$ at high inclinations ($i > 50^{\circ}$). Accounting for azimuthal dependence tightens the $c_4$--$f_{\rm box}$ correlation from $r = -0.79$ to $r = -0.91$.

\section{Properties of the boxy dwarf sample}
\label{app:sample_table}

Table~\ref{tab:boxy_sample} lists the measured properties of the 13 Perseus dwarfs with significantly boxy isophotes ($c_4 < -0.0175$), sorted by $c_4$. Coordinates, magnitudes, central surface brightness ($\mu_{\IE,\rm c}$), S\'ersic index, and axis ratio ($b/a$) are taken from the \citet{EROPerseusDGs} catalog. Effective radii $R_{\rm e}$ are from our pipeline, converted to physical units assuming a Perseus distance of $72$~Mpc. Clustercentric separations $r_{\rm cl}$ are computed relative to NGC~1275.

\begin{table*}
    \centering
    \caption{Measured properties of the 13 boxy Perseus dwarfs, sorted by $c_4$.}
    \label{tab:boxy_sample}
    \footnotesize
    \setlength{\tabcolsep}{4pt}
    \begin{tabular}{lcc rc cc c c c c}
        \hline
        ID & RA (J2000) & Dec (J2000) & $c_4$ & $R_{\rm e}$ & $\IE$ & $\IE - \JE$ & $\mu_{\IE,\rm c}$ & $n_{\rm S\acute{e}rsic}$ & $b/a$ & $r_{\rm cl}$ \\
         & (h:m:s) & (d:m:s) &  & (kpc) &  &  & (mag\,arcsec$^{-2}$) &  &  & (\arcmin) \\
        \hline
        EDwC-0224 & 03:17:48.64 & $+41$:25:19.4 & $-0.050$ & 1.74 & 19.62 & 0.69 & 22.98 & 0.85 & 0.44 & 23.0 \\
        EDwC-0515 & 03:18:46.20 & $+41$:24:17.4 & $-0.041$ & 2.07 & 18.05 & 0.71 & 20.73 & 1.46 & 0.49 & 13.3 \\
        EDwC-0011 & 03:16:18.46 & $+41$:41:12.3 & $-0.032$ & 1.61 & 19.79 & 0.55 & 22.90 & 0.98 & 0.55 & 40.6 \\
        EDwC-0730 & 03:19:24.76 & $+41$:34:21.2 & $-0.028$ & 1.02 & 20.80 & 0.48 & 22.88 & 1.03 & 0.61 & 5.7 \\
        EDwC-0414 & 03:18:27.51 & $+41$:29:47.3 & $-0.025$ & 1.83 & 19.78 & 0.56 & 22.70 & 1.26 & 0.58 & 15.1 \\
        EDwC-0617 & 03:19:03.52 & $+41$:35:13.5 & $-0.024$ & 1.40 & 19.66 & 0.61 & 22.19 & 1.01 & 0.43 & 9.5 \\
        EDwC-1005 & 03:20:27.75 & $+41$:37:25.1 & $-0.023$ & 1.69 & 19.50 & 0.50 & 22.82 & 1.07 & 0.62 & 10.0 \\
        EDwC-0719 & 03:19:22.16 & $+41$:24:26.9 & $-0.022$ & 0.91 & 20.00 & 0.54 & 20.73 & 1.25 & 0.30 & 7.9 \\
        EDwC-0018 & 03:16:24.56 & $+41$:33:42.9 & $-0.022$ & 0.86 & 19.72 & 0.50 & 19.53 & 1.77 & 0.33 & 38.2 \\
        EDwC-0913 & 03:20:07.83 & $+41$:53:37.6 & $-0.020$ & 0.94 & 20.85 & 0.56 & 22.88 & 0.89 & 0.52 & 23.2 \\
        EDwC-0990 & 03:20:24.52 & $+41$:28:45.4 & $-0.019$ & 1.06 & 19.51 & 0.58 & 21.59 & 1.10 & 0.60 & 7.1 \\
        EDwC-0580 & 03:18:57.03 & $+42$:01:33.4 & $-0.019$ & 1.14 & 19.89 & 0.57 & 22.25 & 1.20 & 0.64 & 32.3 \\
        EDwC-0196 & 03:17:42.42 & $+41$:41:28.6 & $-0.018$ & 1.06 & 20.10 & 0.55 & 22.52 & 0.89 & 0.57 & 25.9 \\
        \hline
    \end{tabular}
\end{table*}

\section{Sensitivity of \texorpdfstring{$c_4$}{c4} to centring uncertainty}
\label{app:centre_sensitivity}

\begin{figure*}
    \sidecaption
    \includegraphics[width=12cm]{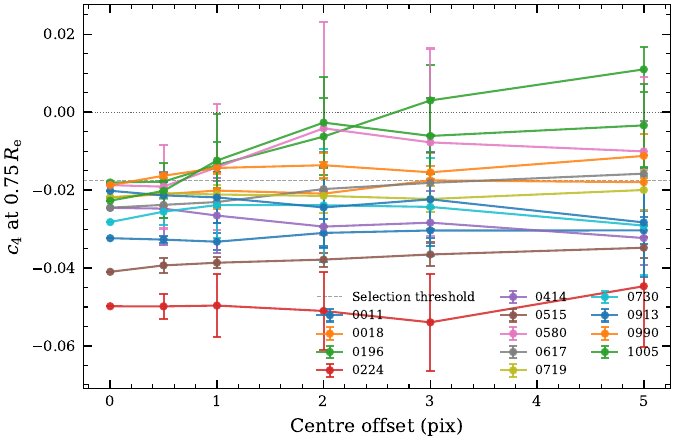}
    \caption{Isophote shape $c_4$ at $0.75\,R_{\rm e}$ versus centre offset for the 13 boxy Perseus dwarfs. Markers show the mean across four cardinal-direction perturbations at the indicated offset; error bars span the standard deviation across them. The dashed grey line marks the boxy selection threshold ($c_4 = -0.0175$).}
    \label{fig:centre_sensitivity}
\end{figure*}

The $c_4$ measurement is calculated relative to a galaxy centre determined by the hill-climbing routine described in Sect.~\ref{sect:center}. To quantify the residual sensitivity of the reported $c_4$ values to centring uncertainty, we re-ran the measurement for each of the 13 boxy dwarfs with the centre artificially perturbed by $\Delta = 0.5, 1, 2, 3$, and 5 pixels in each of the four cardinal directions ($\pm x$, $\pm y$). Figure~\ref{fig:centre_sensitivity} shows the resulting $c_4$ at $0.75\,R_{\rm e}$ as a function of offset; markers are the mean across the four directions and error bars span the standard deviation across them, capturing the directional dependence of $c_4$ at each offset magnitude.

The median $|\Delta c_4|$ is $0.005$ at a 1-pixel offset, $0.008$ at 2 pixels, and $0.009$ at 5 pixels, all substantially smaller than the typical $|c_4| \sim 0.028$ of the boxy sample. The pipeline-determined centres are typically accurate to within $\lesssim 1$ pixel, so the centre-uncertainty contribution to the $c_4$ error budget is comparable to (and somewhat smaller than) the noise-induced uncertainty $\sigma(c_4) \approx 0.003$--$0.008$ established in Appendix~\ref{app:validation}. A few of the marginal galaxies near the selection threshold (EDwC-0196, EDwC-0580, EDwC-1005) cross the threshold only at large ($\geq 2$ pixel) offsets; the remainder are robustly classified as boxy independent of any plausible centring error.

\section{Surface brightness scatter from viewing angle}
\label{app:sb_scatter}

\begin{figure*}
    \sidecaption
    \includegraphics[width=12cm]{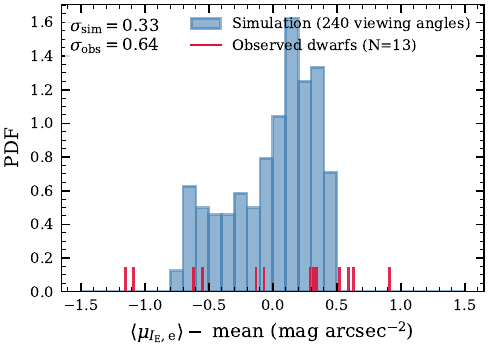}
    \caption{Distribution of mean effective surface brightness $\langle \mu_{\IE,\rm e} \rangle$ relative to the mean, for 240 mock viewing angles of the simulated Med-Rot dwarf at $t = 3.7$~Gyr (blue histogram) and for the 13 observed boxy Perseus dwarfs (red lines).}
    \label{fig:sb_scatter}
\end{figure*}

In Sect.~\ref{sect:interpretation} we note that the projection of a single tidally transformed dwarf at different orientations produces a scatter in the mean effective surface brightness $\langle \mu_{\IE,\rm e} \rangle$ at fixed total luminosity, comparable to the spread observed in our boxy sample. Figure~\ref{fig:sb_scatter} substantiates this claim: the histogram shows the distribution of $\langle \mu_{\IE,\rm e} \rangle$ across 240 mock \textit{Euclid} \IE-band views of the Med-Rot snapshot at $t = 3.7$~Gyr, centred on its mean. Vertical red lines mark the observed values for the 13 boxy Perseus dwarfs, similarly centred on their mean. The simulation's viewing-angle-induced scatter ($\sigma_{\rm sim} = 0.33$~mag~arcsec$^{-2}$, peak-to-peak $\sim 1$~mag~arcsec$^{-2}$) is comparable to but somewhat smaller than the observed sample scatter ($\sigma_{\rm obs} = 0.64$~mag~arcsec$^{-2}$). Because the SKIRT mocks have an arbitrary flux normalisation, only the relative spread is meaningful; both distributions are recentred on zero for comparison.

\end{document}